\PassOptionsToPackage{dvipsnames,table}{xcolor}
\documentclass[journal, ,x11names]{IEEEtran}

\usepackage{setspace}

\usepackage{subfigure}
\usepackage{amsmath,amsfonts}
\usepackage{algorithmic}
\usepackage{algorithm}
\usepackage{array}
\usepackage{subfig}
\usepackage{textcomp}
\usepackage{stfloats}
\usepackage{url}
\usepackage{verbatim}
\usepackage{graphicx}
\usepackage{cite}
\usepackage{subfloat}
\usepackage{tikz}
\usepackage{bm}

\usepackage{color,xcolor}
\usepackage{caption}

\usepackage{pgfplots}
\usetikzlibrary{calc,arrows.meta,positioning}
\usetikzlibrary{shapes.arrows}
\usetikzlibrary{fit,backgrounds} 
\usepackage{standalone}
\usepackage[numbers,sort&compress]{natbib}
\usepackage{setspace}

\begin{document}
\title{Wireless Video Semantic Communication with Decoupled Diffusion Multi-frame Compensation}

\author{Bingyan Xie, Yongpeng Wu,~\IEEEmembership{Senior Member,~IEEE,} Yuxuan Shi, Biqian Feng, Wenjun Zhang,\\~\IEEEmembership{Fellow,~IEEE}, Jihong Park,~\IEEEmembership{Senior Member,~IEEE,} Tony Q.S. Quek,~\IEEEmembership{Fellow,~IEEE}

\thanks{(Corresponding author: Yongpeng Wu)}
\thanks{An earlier version of this paper was submitted to IEEE WCNC 2025. \cite{wvsc}}
\thanks{Bingyan Xie, Yongpeng Wu, Biqian Feng, and Wenjun Zhang are with the Department of Electronic Engineering, Shanghai Jiao Tong University, Shanghai 200240, China (e-mail:bingyanxie, yongpeng.wu, fengbiqian, zhangwenjun@sjtu.edu.cn).}
\thanks{Yuxuan Shi is with the School of Cyber and Engineering, Shanghai Jiao Tong University, Shanghai 200240, China (e-mail:ge49fuy@sjtu.edu.cn).}
\thanks{Jihong Park and Tony Q.S. Quek are with the ISTD Pillar, Singapore University of Technology of Design, 8 Somapah Rd, Singapore 487372 (e-mail:jihong\_park, tonyquek@sutd.edu.sg)}
}

\maketitle
\begin{abstract}
	
Existing wireless video transmission schemes directly conduct video coding in pixel level, while neglecting the inner semantics contained in videos. In this paper, we propose a wireless video semantic communication framework with decoupled diffusion multi-frame compensation (DDMFC), abbreviated as WVSC-D, which integrates the idea of semantic communication into wireless video transmission scenarios. WVSC-D first encodes original video frames as semantic frames and then conducts video coding based on such compact representations, enabling the video coding in semantic level rather than pixel level. Moreover, to further reduce the communication overhead, a reference semantic frame is introduced to substitute motion vectors of each frame in common video coding methods. At the receiver, DDMFC is proposed to generate compensated current semantic frame by a two-stage conditional diffusion process. With both the reference frame transmission and DDMFC frame compensation, the bandwidth efficiency improves with satisfying video transmission performance. Experimental results verify the performance gain of WVSC-D over other DL-based methods e.g. DVSC about 1.8 dB in terms of PSNR.
\end{abstract}


\section{Introduction}

The development of mobile Internet has triggered the prosperity of different video services. Various video-related applications have emerged, e.g. Internet of things, virtual reality, and smart city, which contribute to the major Internet traffic nowadays. To support these applications with huge amount of video data, the efficient wireless video communication technology is on high demand. To this end, traditional wireless video transmission schemes commonly adopt standardized video codecs such as H.264, H.265 \cite{264,265} for source coding, separately followed by the channel codec such as low density parity check (LDPC). Although all digital communication systems today rely on separated source-channel coding (SSCC) schemes, thanks to its modularity, deep learning (DL) based joint source-channel coding (JSCC) is known to achieve higher performance than SSCC schemes in finite blocklength scenarios and to avoid cliff along with the leveling off effects in time-varying channel scenarios \cite{jscc}. With the continuous development of DL technologies, JSCC schemes are expected to be deployed on mobile edge devices or point to point communication scenarios, e.g. exchange of LIDAR or video data among autonomous vehicles for collaborative perception, to mitigate the performance gap between optimal performance brought by the short coding blocklength. These approaches have recently been explored particularly in the domain of semantic communication \cite{cmts,llcm,DeepSC,NTSCC,LCFSC,robust,dvc,dvsc}, which aims to transmit only meaningful information (i.e., semantics) by considering the characteristics of the source data and target tasks against channel perturbations. For example, Jiang et al. \cite{dvc} have proposed a wireless semantic video conferencing system with the incremental redundancy hybrid automatic repeat request to ensure the keypoint transmission in video conference. Niu et al. \cite{dvsc} have proposed an end-to-end DL-enabled video semantic communication system which conducts SNR-adaptive channel coding with semantic restoration to address multi-dimensional noise.

With the above successful attempts for constructing point to point DL-based video semantic communication frameworks, it is promising to introduce the mature standardized video coding structure into the end-to-end video semantic communication framework designs, as \cite{dvsc} does. Similar to the low latency wireless video transmission scenario \cite{dvsc}, for a common Group of Pictures (GoP)-based video coding scheme, the video frame compression is divided into two types. The first frame in a GoP is assumed as the intra frame (I frame) while other frames are predictive frames (P frames). The performance gain of standardized video codecs, like H.264 or H.265, mainly lies in the P frame compression with the inner predictive coding structure which assumes previously reconstructed frame as reference frame and compresses motion vectors and residuals based on the reference frame rather than the whole P frame. In this way, temporal information between adjacent frames is exploited to alleviate the inter-frame redundancy. Actually, a series of emerging deep video coding schemes \cite{DVC, MLVC, ALVC} for inter-frame coding are inspired by traditional predictive coding structures to build DL-based video compression networks. As summarized in Fig. 1(a), they directly conduct deep video coding in pixel level and collect both motion vectors and residuals of each P frame, as traditional codec does. Meanwhile, since transmission cost is not directly considered in video coding, the predicted frames are shared across the transceiver to help recover original video frames. However, for the wireless video transmission, the predicted frame is obviously unknown at the receiver, which would bring extra communication overhead for the video transmission. To tackle this, \cite{dvsc} assumes previously reconstructed video frames as background knowledge and thus conducts motion compensation at the receiver based on such frames. As shown in Fig. 1(b), although it avoids the transmission of predicted frames, motion vectors and residuals are still transmitted. Moreover, such pixel-level wireless video transmission structure heavily relies on the accuracy of received motion information, bringing performance degradation when faced with serious channel impairment. Therefore, we propose a semantic-level wireless video transmission structure in Fig. 1(c), as the previous work does \cite{wvsc}. Inspired by the feature-level coding scheme \cite{fvc}, DL-based video coding is conducted in the semantic level to further alleviate the temporal redundancy from original successive video signals. The encoded video frames refer to the semantic frames. Unlike the pixel-level wireless video communication \cite{dvsc}, the first extracted frame in a GoP is utilized as the reference semantic frame all the way for the transmission of every semantic P frame. The reference frame refers to semantic I frame, which can be transmitted independently similar to common DL-based image transmission. While for semantic P frames, residuals between each semantic P frame and semantic I frame are computed through motion estimation and compensation. In this way, only the semantic I frame and residuals of each semantic P frame are transmitted, greatly improving the communication efficiency.

\begin{figure}[htbp]
	\centering
	\includegraphics[width=3.3in]{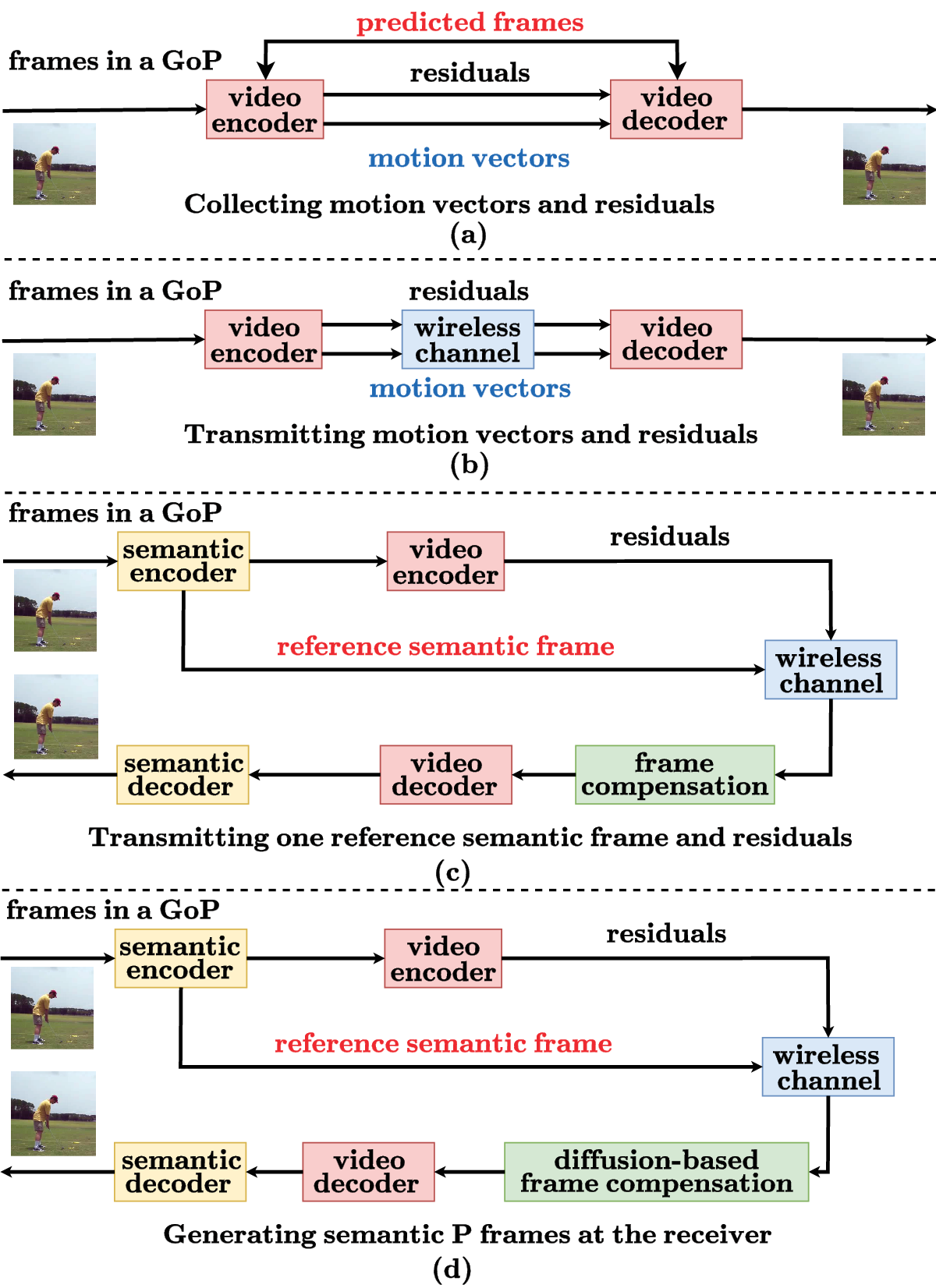}
	\caption{A comparison of video compression and wireless video communication methods: (a) pixel-level video compression without communication \cite{DVC, MLVC, ALVC}. (b) pixel-level wireless video communication \cite{dvsc}. (c) semantic-level wireless video communication. \cite{wvsc} (d) proposed diffusion-based semantic-level wireless video communication.}
	\label{fig_1}
\end{figure}

Although the semantic-level wireless video transmission provides an insightful structure for the communication-efficient video semantic communication framework, the absent of motion vectors would pose much difficulty for the reconstruction of successive semantic frames at the receiver. Temporal information within the motion vectors are essential for the video decoder to conduct motion compensation. \cite{dvc} utilizes the I frame as shared semantic knowledge and transmits only keypoints for P frame restoration. \cite{gfvc} collects the reconstructed key-reference picture coded by traditional video codec and recontruct subsequent pictures based on such key pictures for generative face video coding. However, these above straightforward video frame inference method is only applied for stable scenarios, e.g. wireless video conferencing with simple facial changes, which would face serious performance degradation under diverse video contents. Fortunately, previously reconstructed semantic frames contain adequate motion information required for video decoding. To retain the robustness of semantic-level video decoding, it is of urgent requirement to compensate the motion differences between received semantic I frame and each semantic P frame at the receiver aided by previously reconstructed semantic frames, as the `frame compensation' module in Fig. \ref{fig_1}(c). Specifically, the received semantic I frame is essential to be steered to the direction of corresponding semantic P frame for the latter motion estimation and compensation process, which is typically a conditional generation problem. Diffusion-based models, which are widely used in artificial intelligence generated content \cite{cft,cdd,360dvd,diffeditor}, work as promising solutions to generate the required semantic frame at the receiving end with multi-frame conditions. Diffusion models learn the feature distribution through forward noise estimation and generate conditional samples through iterative reverse sampling, which can be much more powerful for generating high-quality samples compared to existing visual feature extraction models \cite{vit,cnn} and generative models \cite{gan, vae}. Existing works \cite{dmce,cddm} have utilized the strong generation ability of diffusion-based models to conduct signal processing in wireless communications. Zeng et al. \cite{dmce} have considered multi-user scenario and thus utilized the diffusion model to learn the particular data distribution of channel effects on the transmitted semantic features. Wu et al. \cite{cddm} have found the correlation between diffusion noise forward process and the wireless channel transmission. Channel denoising diffusion models (CDDM) have been proposed which conduct denoising diffusion probabilistic model (DDPM) \cite{ddpm} reverse sampling for signal denoising at the receiver. Such successful implementation implies great potential for employing diffusion-based models for multi-frame compensation.

\begin{table}[htbp]
	\centering
	\caption{Limitations of existing works.}
	\label{table1}
	
	\begin{tabular}{|c|c|}  
		\hline 
		&\\[-6pt] 
		Framework&Limitations \\
		\hline
		&\\[-6pt]  
		WVSC \cite{wvsc}&Hard to retain temporal consistency\\
		\hline
		&\\[-6pt] 
		DVC \cite{dvc}&Limited to wireless video conferencing\\
		\hline
		&\\[-6pt] 
		DVSC \cite{dvsc}&Less efficient pixel-level video transmission\\
		\hline
	\end{tabular}
\end{table}

Based on the above limitations of existing wireless video transmission framework in Tab. \ref{table1}, we thus provided the wireless video semantic communication framework with decoupled diffusion multi-frame compensation, termed as WVSC-D. As depicted in Fig. \ref{fig_1}(d), semantic P frames are generated at the receiver through diffusion models. To exploit the temporal information embedded in previously reconstructed frames, the key challenge is how to integrate such multi-frame conditions into diffusion process for compensating current semantic P frames with relatively affordable computation cost. CDDM enables signal denoising under wireless channels but with lots of sampling steps, which is not practical for the wireless video transmission. To fully utilize the strong generative ability of diffusion models with few sampling steps, we focus on the P frame generation and thus propose a generation-based multi-frame compensation (GMFC) strategy which modulates the sampling trajectory of the diffusion process to steer the compensation result to the direction of current semantic P frame step by step. Based on the GMFC, to ensure the temporal consistency of generated semantic frames, we further propose the decoupled diffusion multi-frame compensation (DDMFC) module. DDMFC decouples both frame and generated noise into two parts, which are base component and residual component. For the frame decomposition, each input frame for the DDMFC is the combination of both received semantic I frame and corresponding residuals for each semantic P frame. While for the noise decomposition, the base noise is generated by CDDM, which entirely relies on the received semantic I frame. The unique residual noises for each semantic P frame are then learnt based on the shared base frame and noise. Through both base components and residual components, the polished current semantic P frames can be computed. Since the generated semantic P frame is composed of shared common part of received semantic I frame and private part guided by multi-frame conditions, both temporal consistency and frame uniqueness for the generated semantic P frames are protected during multi-frame compensation.

The contributions of this paper are summarized as follows  
\begin{enumerate}
	\item{}
	To enhance bitrate saving by reducing redundant I and P frames, we propose a novel wireless video transmission framework, coined WVSC-D. The main idea of WVSC-D is to conduct DL-based video coding on extracted semantics rather than directly on pure video signals. Instead of transmitting motion vectors for every semantic frame, the reference frame is transmitted only once in a GoP. Only key semantic frame and residual information are transmitted, greatly reducing the communication overhead.
	\item{}
	To compensate semantic P frames at the receiver in WVSC-D, we introduce a GMFC strategy. In GMFC, the rich temporal information is exploited by polishing the received semantic I frame towards the direction of current semantic P frame guided by multi-frame conditions. For each semantic P frame at the receiver, energy-based loss functions are introduced to convert multi-frame conditions into explicit gradient values during each reverse sampling step. Such gradient values provide direct guidance and help steer the received semantic I frame to the direction of each current semantic P frame step by step. 
	\item{}
	To improve temporal consistency of semantic P frames, we develop a DDMFC module based on the GMFC strategy. In DDMFC, the whole diffusion process can be decoupled into two parts, including base component generation and residual noise generation. The received semantic I frame and estimated noises in the base noise generation model are assumed as basic components and shared across each generation process for other semantic P frames in the same GoP. For the semantic P frames, unique residual noise is learned to generate each semantic P frame separately.
	\item{}
	To generate residual noise for semantic P frames, we additionally integrate a multi-frame fusion attention (MFA) structure into WVSC-D. It helps the diffusion network generate the residual noise aware of temporal information. The previously reconstructed frames and the current diffusion samples are fused together through the cross attention mechanism. These fused samples serve as input for the residual noise generation.
\end{enumerate}

Note that the conference version \cite{wvsc} of this work presents a wireless video semantic communication (WVSC) framework without GMFC and DDMFC, hereafter termed WVSC. In contrast to WVSC-D, though WVSC meets the framework structure in Fig. \ref{fig_1}(c), it only applies the MFA to conduct the multi-frame compensation at the receiver, so struggling with temporal inconsistency in reconstructed frames. In this paper, to fully exploit the temporal information embedded in previously reconstructed frames, a series of generation-based diffsuion compensation schemes and algorithms are proposed, as depicted in Fig. \ref{fig_1}(d). The superior performance of WVSC-D against other traditional or DL-based schemes will be demonstrated by simulation in Sec. V.

The rest of this paper is organized as follows. Section II introduces the system model and the proposed framework of WVSC-D. Section III illustrates the framework details, including deep video coding in semantic level, generation-based multi-frame compensation and decoupled diffusion process. Section IV describes the deployment of WVSC-D including network structures, training loss and strategies. Section V demonstrates the superiority of the proposed networks through a series of experiments. Section VI concludes the paper.

Notations: $\mathbb{R}$ and $\mathbb{C}$ refer to the real and complex number sets, respectively. $\mathcal{CN}\left (\mu, \sigma^2 \right)$ denotes a complex Gaussian distribution with mean $\mu$ and variance $\sigma^2$. $\nabla_\mathbf{x}(\cdot)$ denotes the gradient w.r.t $\mathbf{x}$. $\mathrm{diag}(\cdot)$ refer to the diagonalization operations between a vector and its corresponding diagonal matrix. $|\cdot|$ refers to computing the modulus of a complex number. $\ast$ refers to the element-wise multiplication. $\mathbf{I}$ denotes the unit matrix. The operator $\left(\cdot\right)^{T}$ denotes the matrix transpose. $D_{\mathrm{KL}}(q(\cdot)||p(\cdot))$ refers to the Kullback-Leibler (KL) divergence between two distributions. $\text{Re}(\cdot)$, $\text{Im}(\cdot)$ refer to the real part and imaginary part of the complex value, respectively.

\begin{figure*}[htbp]
	\centering
	\includegraphics[width=6.8in]{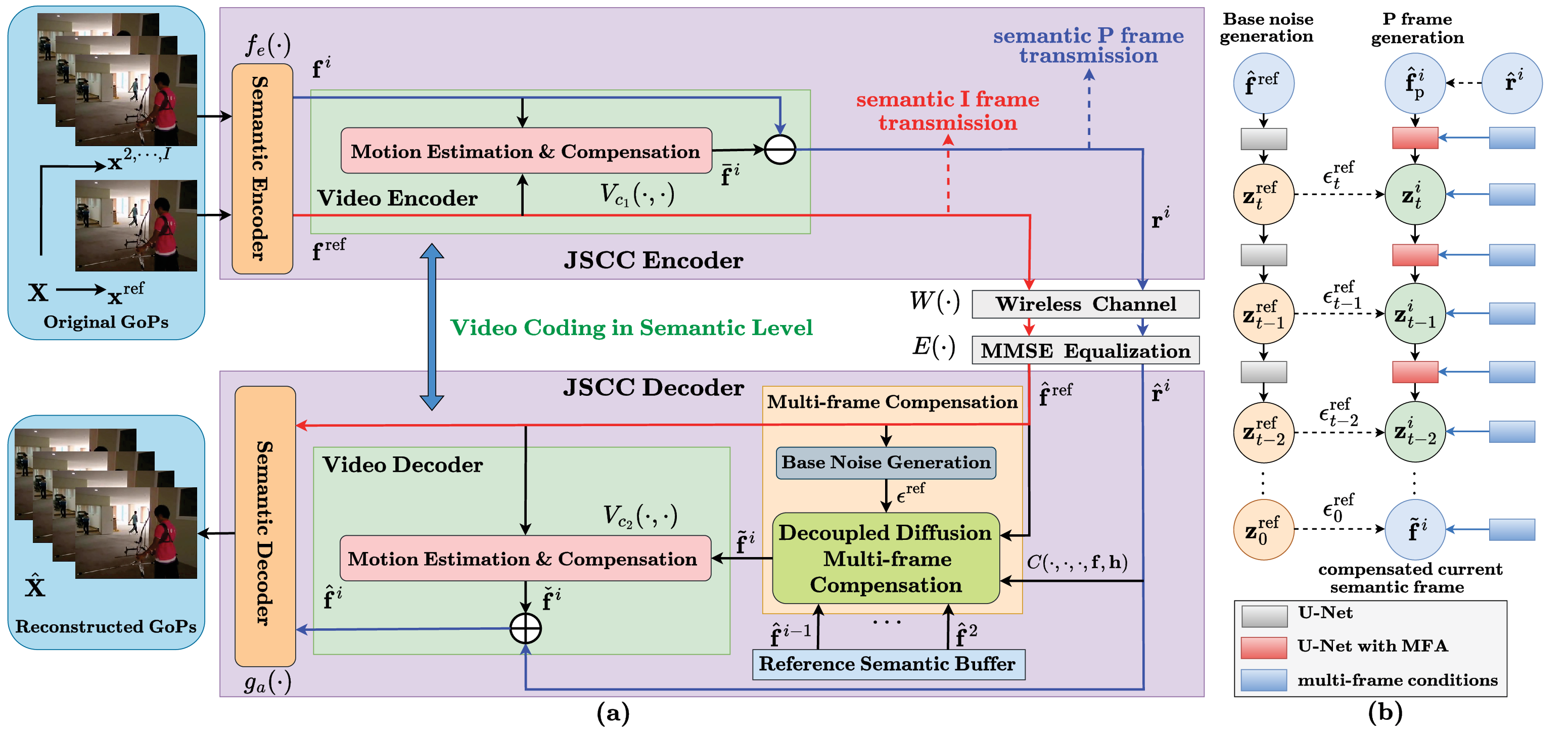}
	\caption{(a) The proposed WVSC-D framework. The video is transmitted by a series of GoPs, which is divided into semantic I frame transmission (red lines) and semantic P frame transmission (blue lines), respectively. (b) The proposed DDMFC module. The received semantic I frames $\hat{\mathbf{f}}^\mathrm{ref}$ are compensated into current semantic P frames $\tilde{\mathbf{f}}^i$ through conditional diffusion generation.}
	\label{fig_2}
\end{figure*}

\section{System model and proposed framework}
In this section, we describe the system model of WVSC-D framework by both semantic I frame transmission and semantic P frame transmission following the low latency scenarios (IPPP mode).

\subsection{Semantic I frame transmission}
We consider a GoP-based wireless video delivery system. Given a video sequence $\mathcal{X}= \left \{\mathbf{X}_1,\mathbf{X}_2,\cdots,\mathbf{X}_N \right\}$, where $\mathbf{X}_n\in\mathbb{R}^{I\times H\times W\times3}, n=1,\cdots, N$, which is composed of $N$ successive GoPs. For an arbitrary GoP $\mathbf{X} = \left\{\mathbf{x}^1,\mathbf{x}^2,\cdots,\mathbf{x}^I\right\}\in\mathcal{X}$, where $I$ is the total number of frames in a GoP, $\mathbf{x}^i\in \mathbb{R}^{H\times W\times3}$, we treat the first frame $\mathbf{x}^1$ as the reference frame $\mathbf{x}^{\mathrm{ref}}$. The reference frame, which is also the I frame in a GoP, is communicated using DL-based JSCC, i.e., the semantic encoder and decoder in Fig. \ref{fig_2}(a), similar to the simple image semantic communications. The semantic encoder, $f_e(\cdot): \mathbb{R}^{I\times H\times W\times 3}\longrightarrow \mathbb{R}^{I\times L}$, encodes $\mathcal{X}$ into the semantic sequence, $\mathcal{F}=\left\{\mathbf{f}^1,\mathbf{f}^2,\cdots,\mathbf{f}^I\right\}$, where $L$ is the code length for the extracted semantic frame with $\mathbf{f}^i\in \mathbb{R}^{L}$. Here $\mathbf{f}^{\mathrm{ref}}=\mathbf{f}^1$ is the semantic I frame. Then, the semantic I frame is transmitted through wireless channels and equalized by the minimum mean square error (MMSE) channel equalization after conversion between real and complex values, which can be formulated as
\begin{align}
	\hat{\mathbf{f}}^{\mathrm{ref}}=\mathbf{H}_s\ast\mathbf{f}^{\mathrm{ref}}+\mathbf{H}_n\ast\mathbf{n}, 
\end{align}
where $\mathbf{H}_s=\mathrm{diag}(\mathbf{h}_d^2(\mathbf{h}_d^2+2\sigma^{2}\mathbf{I})^{-1})\in\mathbb{R}^{L}$ along with $\mathbf{H}_n=\mathrm{diag}(\mathbf{h}_d(\mathbf{h}_d^2+2\sigma^{2}\mathbf{I})^{-1})\in\mathbb{R}^{L}$ refers to the channel equalization parameters, $\mathbf{h}_d = \mathrm{diag}(\begin{bmatrix}
	|\mathbf{h}|\\
	|\mathbf{h}|
\end{bmatrix})\in \mathbb{R}^{L\times L}$, $\mathbf{h}\in \mathbb{C}^{L/2}$ denotes the Rayleigh channel fading index following the distribution of $\mathcal{CN}(0,1)$, $\mathbf{n}=\begin{bmatrix}
\text{Re}(\mathbf{n}_c)\\
\text{Im}(\mathbf{n}_c)
\end{bmatrix}\in \mathbb{R}^{L}$, $\mathbf{n}_c\in \mathbb{C}^{L/2}$ is the complex Gaussian channel noise following the distribution of $\mathcal{CN}(0,2\sigma^{2})$.

At the receiver, the semantic decoder, $g_a(\cdot): \mathbb{R}^{I\times L}\longrightarrow \mathbb{R}^{I\times H\times W\times 3}$, converts $\hat{\mathcal{F}}=\left\{\hat{\mathbf{f}^1},\hat{\mathbf{f}^2},\cdots,\hat{\mathbf{f}^I}\right\}$ into the final reconstructed video sequence $\hat{\mathcal{X}}=\left\{\hat{\mathbf{X}}^1,\hat{\mathbf{X}}^2,\cdots,\hat{\mathbf{X}}^I\right\}$. The whole transmission process for the semantic I frame can be formulated as
\begin{align}
	\mathbf{x}^{\mathrm{ref}} \xrightarrow{f_e(\cdot)} \mathbf{f}^{\mathrm{ref}} \xrightarrow{E\left(W(\cdot)\right)} \hat{\mathbf{f}}^{\mathrm{ref}} \xrightarrow{g_a(\cdot)} \mathbf{\hat{x}}^{\mathrm{ref}},
\end{align}
where $W(\cdot)$ implies wireless channels and $E(\cdot)$ refers to the MMSE channel equalization.

\subsection{Semantic P frame transmission}
For the semantic P frame transmission, non-reference semantics $\mathbf{f}^i$ with $i\geq 2$, are communicated using not only DL-based JSCC but also additional pre-/post-processing. They are processed by the motion estimation $\&$ compensation video encoder, $V_{c_1}(\cdot, \mathbf{f}^{\mathrm{ref}}): \mathbb{R}^{L}\longrightarrow \mathbb{R}^{L}$, along with the reference semantics $\mathbf{f}^{\mathrm{ref}}$ to acquire the predicted semantic frame $\bar{\mathbf{f}}^i\in \mathbb{R}^{L}$. Then, the residual $\mathbf{r}^i\in \mathbb{R}^{L}$ can be computed through the subtraction between $\mathbf{f}^i$ and $\bar{\mathbf{f}}^i\in \mathbb{R}^{L}$. In this way, the transmission for the P frame semantics are substituted by the corresponding sparse residuals which is much easier to be compressed than original video frames. Notably, in common video coding schemes, $\bar{\mathbf{f}}^i$ is known for both compression and decompression process. However, for wireless video delivery scenarios, $\bar{\mathbf{f}}^i$ is obviously unknown at the receiving end. Instead of transmitting motion vectors for every semantic P frame to recover $\bar{\mathbf{f}}^i$, we utilize the consistent reference semantic frame $\mathbf{f}^{\mathrm{ref}}$ as the substitute for the motion vectors, greatly saving the communication cost. Since residuals of different semantic P frames can be transmitted in parallel and the transmission interval between I frame and P frame is quite small, we assume the received codewords within one specific GoP undergo the same channel fading. The wireless transmission process for the residuals of semantic P frames can be formulated as
\begin{align}
	\hat{\mathbf{r}}^i=\mathbf{H}_{s}\ast\mathbf{r}^i+\mathbf{H}_{n}\ast\mathbf{n}_\mathbf{r}^i, 
\end{align}
where $\mathbf{n}_\mathbf{r}^i=\begin{bmatrix}
	\text{Re}(\mathbf{n}_\mathbf{rc}^i)\\
	\text{Im}(\mathbf{n}_\mathbf{rc}^i)
\end{bmatrix}\in \mathbb{R}^{L}$, $\mathbf{n}_\mathbf{rc}^i\in \mathbb{C}^{L/2}$ is the complex Gaussian channel noise following the distribution of $\mathcal{CN}(0,2\sigma^{2})$.

At the receiver, with the received $\hat{\mathbf{f}}^{\mathrm{ref}}$ and $\hat{\mathbf{r}}^i$, we obtain the reconstructed semantic sequence, $\hat{\mathcal{F}}=\left\{\hat{\mathbf{f}}^1,\hat{\mathbf{f}}^2,\cdots,\hat{\mathbf{f}}^I\right\}$, using a series of frame compensation and video decoding operations. To conduct multi-frame compensation at the receiver, the base noise generation stage utilizes CDDM to produce the estimated noise $\mathbf{\epsilon}^\mathrm{ref}\in \mathbb{R}^{L}$ as guidance for the latter frame generation. Based on the received semantic I frame $\mathbf{\hat{f}}^{\mathrm{ref}}$, base noise $\mathbf{\epsilon}^\mathrm{ref}$ and the corresponding residuals $\mathbf{\hat{r}}^i$, the decoupled diffusion multi-frame compensation module, $C(\cdot,\cdot,\cdot,\mathbf{f}, \mathbf{h}): \mathbb{R}^{L}\times\mathbb{R}^{L}\times\mathbb{R}^{L}\longrightarrow \mathbb{R}^{L}$, is employed to polish $\mathbf{\hat{f}}^{\mathrm{ref}}$ into the current semantic P frame $\mathbf{f}^i$ as much as possible, where $\mathbf{f}=\left\{\mathbf{\hat{f}}^{i-1},\cdots,\mathbf{\hat{f}}^{2}\right\}$ refers to the previous multiple reconstructed frames. $\mathbf{\tilde{f}}^i \in\mathbb{R}^{L}$ refers to the polished current semantic frame after multi-frame compensation. As shown in Fig. \ref{fig_2}(b), multi-frame conditions are employed to generate semantic P frames with the challenges of both channel impairment and motion differences among non-adjacent frames. The compensation module is empowered by the conditional diffusion, in which $\mathbf{\tilde{f}}^i$ is generated by the reverse sampling process. The MFA module is embedded in the diffusion network for semantic P frame generation. After multi-frame compensation, through motion estimation $\&$ compensation video decoder $V_{c_2}(\cdot, \mathbf{\hat{f}}^{\mathrm{ref}}): \mathbb{R}^{L}\longrightarrow \mathbb{R}^{L}$, the reconstructed predicted semantic frame $\mathbf{\check{f}}^i\in\mathbb{R}^{L}$ can be acquired. With $\hat{\mathbf{r}}^i$ and $\mathbf{\check{f}}^i$, we are able to acquire the current reconstructed semantic frames $\mathbf{\hat{f}}^i$ by adding them together. Finally, the semantic decoder decodes $\mathbf{\hat{f}}^i$ into final reconstructed video frames $\mathbf{\hat{x}}^i$. The whole process can be formulated as
\begin{align}
	\mathbf{x}^i \xrightarrow{f_e(\cdot)} \mathbf{f}^i \xrightarrow{V_{c_1}(\cdot, \mathbf{f}^{\mathrm{ref}})} \mathbf{\bar{f}}^i \xrightarrow{-\mathbf{f}^{\mathrm{ref}}} \mathbf{r}^i, i\ne1
\end{align}
\begin{align}
	\mathbf{\hat{f}}^{\mathrm{ref}} \xrightarrow{C(\cdot,\cdot,\cdot,\mathbf{f}, \mathbf{h})} \mathbf{\tilde{f}}^i \xrightarrow{V_{c_2}(\cdot, \mathbf{\hat{f}}^{\mathrm{ref}})} \mathbf{\check{f}}^i \xrightarrow{+\hat{r}^i} \mathbf{\hat{f}}^i \xrightarrow{g_a(\cdot)} \mathbf{\hat{x}}^i,i\ne1
\end{align}

In this way, spatial-temporal redundancy is significantly reduced at the semantic level, which provides a far more compact and efficient feature space for compression compared to the traditional or existing pixel-level domain. This semantic representation, by distilling the fundamental meaning of a scene into a structured set of features, inherently eliminates superfluous pixel-level details that are costly to transmit. Conversely, conventional pixel-space operations are frequently hampered by challenges such as inaccurate motion estimation and ineffective motion compensation, particularly in complex dynamic scenes, leading to artifacts and suboptimal compression rates \cite{fvc}. By integrating the robustness of semantic communication with the representational power of deep video coding, this fusion enables a highly efficient and resilient framework for wireless video transmission, ensuring superior performance even under bandwidth constraints and unreliable channel conditions.

\section{Details of WVSC-D Framework}
In this section, we present the structure and detailed designs of each module in the WVSC-D framework.

\subsection{Deep Video Coding in Semantic Level}
\begin{figure}[htbp]
	\centering
	\includegraphics[width=3.5in]{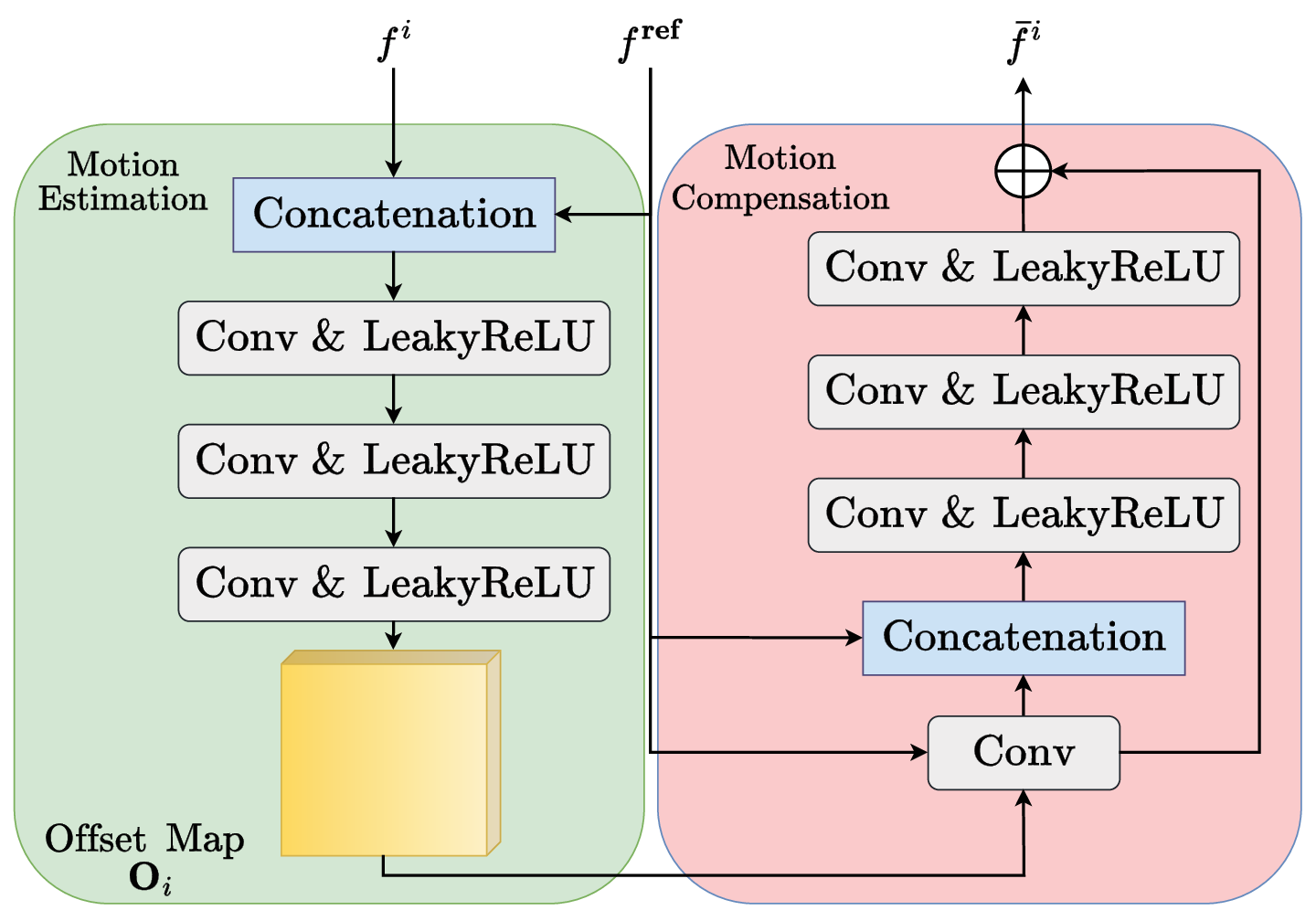}
	\caption{The structure of motion estimation $\&$ compensation network.}
	\label{fig_3}
\end{figure}
As illustrated in Section \uppercase\expandafter{\romannumeral2}, WVSC-D conducts deep video coding in semantic level. The inter-frame correlation is exploited based on extracted semantic frames rather than directly on original frames. The structure of motion estimation $\&$ compensation video coder is shown in Fig. \ref{fig_3}. It is mentioned that the video encoder ${V_{c_1}(\cdot, \mathbf{f}^{\mathrm{ref}})}$ and decoder $V_{c_2}(\cdot, \mathbf{\hat{f}}^{\mathrm{ref}})$ share the same network structure. The video coder is composed of two parts: motion estimation and motion compensation. For the motion estimation, the current semantic P frame $\mathbf{f}^i$ and the semantic I frame $\mathbf{f}^{\mathrm{ref}}$ are concatenated together as input. Then through a lightweight structure composed of a series of convolutional layers and activation layers with LeakyReLU \cite{leakyrelu}, the motion offset map $\mathbf{O}_i \in\mathbb{R}^{L}$ between reference and current semantics is generated. In other words, $\mathbf{O}_i$ refers to the motion vector in common video coding frameworks. For the motion compensation, with the estimated offset map, the predicted semantic frame $\mathbf{\bar{f}}^i$ can be produced by the motion compensation network which is based on the residual convolutional networks.

\subsection{Generation-based Multi-frame Compensation}
Although WVSC-D structure greatly saves communication cost in wireless video transmission, the motion differences between non-adjacent frames hinder its performance. To be specific, transmitting only semantic I frame and compressed residuals brings extra difficulty to the video decoding of successive semantic frames. Based on the key reference semantic I frame, a compensation strategy for steering semantic I frame towards the direction of each semantic P frame aided by multi-frame conditions is necessary at the receiver, which is named as GMFC. As provided in \cite{cddm}, the reverse sampling process can be utilized to denoise the received signals in specific cases. However, such case is only met when the number of sampling steps is large, which incurs much computation cost for the wireless video transmission once utilizing CDDM for denoising semantic I frame. Fortunately, as illustrated in \cite{ddim}, the strong generation ability of diffusion models allows relatively satisfying sample production with limited sampling steps. Specifically, about 50$\times$ or even 100$\times$ sampling speed up compared to DDPM \cite{ddpm} can grasp reasonable high-level features. To achieve the trade-off between video transmission performance and overall computation cost, the base noise generation module with CDDM is employed to produce base noise for the semantic P frame compensation rather than signal denoising for the semantic I frame. For the current semantic P frame, GMFC introduces multi-frame conditions to generate the estimated current frame for alleviating the channel noise and motion differences among frames.

For the semantic I frame $(i=1)$, the sample is produced according to CDDM \cite{cddm} as
\begin{equation}
	\begin{aligned}
		\mathbf{z}_{t-1}^\mathrm{ref}&=\sqrt{\bar{a}_{t-1}}( \frac{\mathbf{z}_{t}^\mathrm{ref}-\sqrt{1-\bar{a}_t}\mathbf{H}_n\ast\mathbf{\epsilon}_t^\mathrm{ref}(\mathbf{z}_{t}^\mathrm{ref})}{\sqrt{\bar{a}_t}})\\&+\sqrt{1-\bar{a}_{t-1}-\sigma^2_t} \mathbf{H}_n\ast\mathbf{\epsilon}_t^\mathrm{ref}(\mathbf{z}_{t}^\mathrm{ref})+\sigma_t\mathbf{\epsilon} 
	\end{aligned}
\end{equation}
where $\mathbf{z}_{t}^\mathrm{ref}$ refers to the output state of diffusion at a specific sampling timestep $t$, $\mathbf{\epsilon}_t^\mathrm{ref}(\mathbf{z}_{t}^\mathrm{ref})$ is the estimated noise of diffusion, $\sigma_t$ is the corresponding variance of the reverse step which controls the diversity of the generated samples, $\epsilon \sim \mathcal{N}(0,1)$ is the added Gaussian noise, $\bar{a}_{t}={\textstyle\prod_{i=1}^{t}}a_i$ and $a_i\in (0,1)$ are hyperparameters.

Here we provide a concise review in terms of the difference between common DDPM and CDDM. The diffusion process of the semantic I frame is presented in Fig. \ref{fig_4}(a). While both methods employ the same identical forward process that incrementally adds Gaussian noise to a semantic I frame over T timesteps, their reverse processes differ significantly. In common DDPM, the reverse process starts from the $T$-th timestep with lots of sampling step to recover the original images. CDDM innovatively aligns the forward diffusion with wireless channel transmission, allowing the reverse process to commence from an intermediate timestep $m$ that corresponds to the equivalent noise level during transmission, thereby substantially reducing computational cost by sampling from this earlier point to generate the refined semantic frame.

As opposed to the diffusion denoising for the semantic I frame, directly generating current semantic P frame from received semantic I frame requires sufficient temporal information between such two adjacent frames. So the multiple previous reconstructed frames serve as the complement of temporal information for the conditional diffusion generation. To exploit the multi-frame conditions, the energy-based model (EBM) \cite{ebm} is introduced to formulate the diffusion process, which can be defined as
\begin{align}
	p(\mathbf{z})=\frac{\mathrm{exp}(-V(\mathbf{z}))}{P}, 
\end{align}
where $V(\mathbf{z})$ denotes the corresponding energy function across states $\mathbf{z}$, $P$ denotes a normalization constant.

Through Langevin equation \cite{langevin}, the diffusion reverse sampling step is described as
\begin{align}
	\mathbf{z}_{t-1}^i=\mathbf{z}_t^i-\nabla_{\mathbf{z}_t^i}\mathrm{log}p(\mathbf{z}_{t-1}^i|\mathbf{z}_t^i)+\mathbf{\epsilon},\mathbf{\epsilon} \sim \mathcal{N}(0,I).
\end{align}
where $\nabla_{\mathbf{z}_t^i}\mathrm{log}p(\mathbf{z}_{t-1}^i|\mathbf{z}_t^i)$ refers to the score function of the density $p(\mathbf{z}_{t-1}^i|\mathbf{z}_t^i)$.

With the above EBM formulation, it is convenient to modulate such energy functions to satisfy given criteria, e.g. different conditions. Similar to \cite{steer}, thus we add the multi-frame conditions and formulate the conditional transition probability $p(\mathbf{z}_{t-1}^i|\mathbf{z}_t^i,\mathbf{f}, \mathbf{h})$ as
\begin{align}
	p(\mathbf{z}_{t-1}^i|\mathbf{z}_t^i,\mathbf{f}, \mathbf{h})\propto \frac{p(\mathbf{z}_{t-1}^i|\mathbf{z}_t^i)p(\mathbf{f}, \mathbf{h}|\mathbf{z}_{t-1}^i)}{p(\mathbf{f}, \mathbf{h}|\mathbf{z}_t^i)}. 
\end{align}

Using the log of probability density in Eq. (7), the effective score for conditional transition can be represented as
\begin{equation}
	\begin{aligned}
		\nabla_{\mathbf{z}_t^i}\mathrm{log}p(\mathbf{z}_{t-1}^i|\mathbf{z}_t^i,\mathbf{f}, \mathbf{h})&=\nabla_{\mathbf{z}_t^i}\mathrm{log}p(\mathbf{z}_{t-1}^i|\mathbf{z}_t^i)\\&-\nabla_{\mathbf{z}_t^i}V_1(\mathbf{z}_t^i,\mathbf{f}, \mathbf{h})+\nabla_{\mathbf{z}_t^i}V_2(\mathbf{z}_{t-1}^i,\mathbf{f}, \mathbf{h}),
	\end{aligned}
\end{equation}
where $\nabla_{\mathbf{z}_t^i}\mathrm{log}p(\mathbf{z}_{t-1}^i|\mathbf{z}_t^i)$ denotes the aforementioned unconditional diffusion sampling step as in Eq. (6). $V_1(\cdot)$ and $V_2(\cdot)$ denote the corresponding energy loss functions that model the conditional distributions of $\mathbf{z}_t^i$ and $\mathbf{z}_{t-1}^i$ given the previously reconstructed frames $\mathbf{f}$ and channel fading $\mathbf{h}$, respectively. 

Finally, similar to the unconditional case Eq. (6), the conditional sample can be generated as
\begin{equation}
	\begin{aligned}
		\mathbf{z}_{t-1}^i&=\sqrt{\bar{a}_{t-1}}( \frac{\mathbf{z}_{t}^i-\sqrt{1-\bar{a}_t}\mathbf{H}_n\ast\mathbf{\epsilon}_t^\mathrm{ref}(\mathbf{z}_{t}^i)}{\sqrt{\bar{a}_t}})\\&+\sqrt{1-\bar{a}_{t-1}-\sigma^2_t} \mathbf{H}_n\ast\mathbf{\epsilon}_t^\mathrm{ref}(\mathbf{z}_{t}^i)+\sigma_t\mathbf{\epsilon} \\&-k(t)\underbrace{\nabla _{\mathbf{z}_t^i}(V_1(\mathbf{z}_t^i,\mathbf{f}, \mathbf{h})-V_2(\mathbf{z}_{t-1}^i,\mathbf{f}, \mathbf{h}))}_{\text{multi-frame conditional steering}},  
	\end{aligned}
\end{equation}
where $k(t)\in \mathbb{R}^{1}$ refers to the scaling factor which controls the extent of multi-frame conditional steering.

From Eq. (11), the current semantic P frames are generated by the iterative sampling steps which are composed of unconditional sampling along with the condition steering formulated by the gradient values of energy loss functions. Note that the unconditional reverse sampling step is the same as the CDDM reverse denoising step. In other words, semantic P frames are produced based on the estimated noise of the reference semantic frame. Moreover, since the EBM allows the energy modulation to the most necessary direction, the multi-frame conditions work as guidance for each sampling step to progressively steer to the direction of current semantic P frame.

\subsection{Decoupled the Diffusion Process with the Correlation between semantic I frame and Current semantic P frame}
\begin{figure*}[htbp]
	\centering
	\includegraphics[width=6.8in]{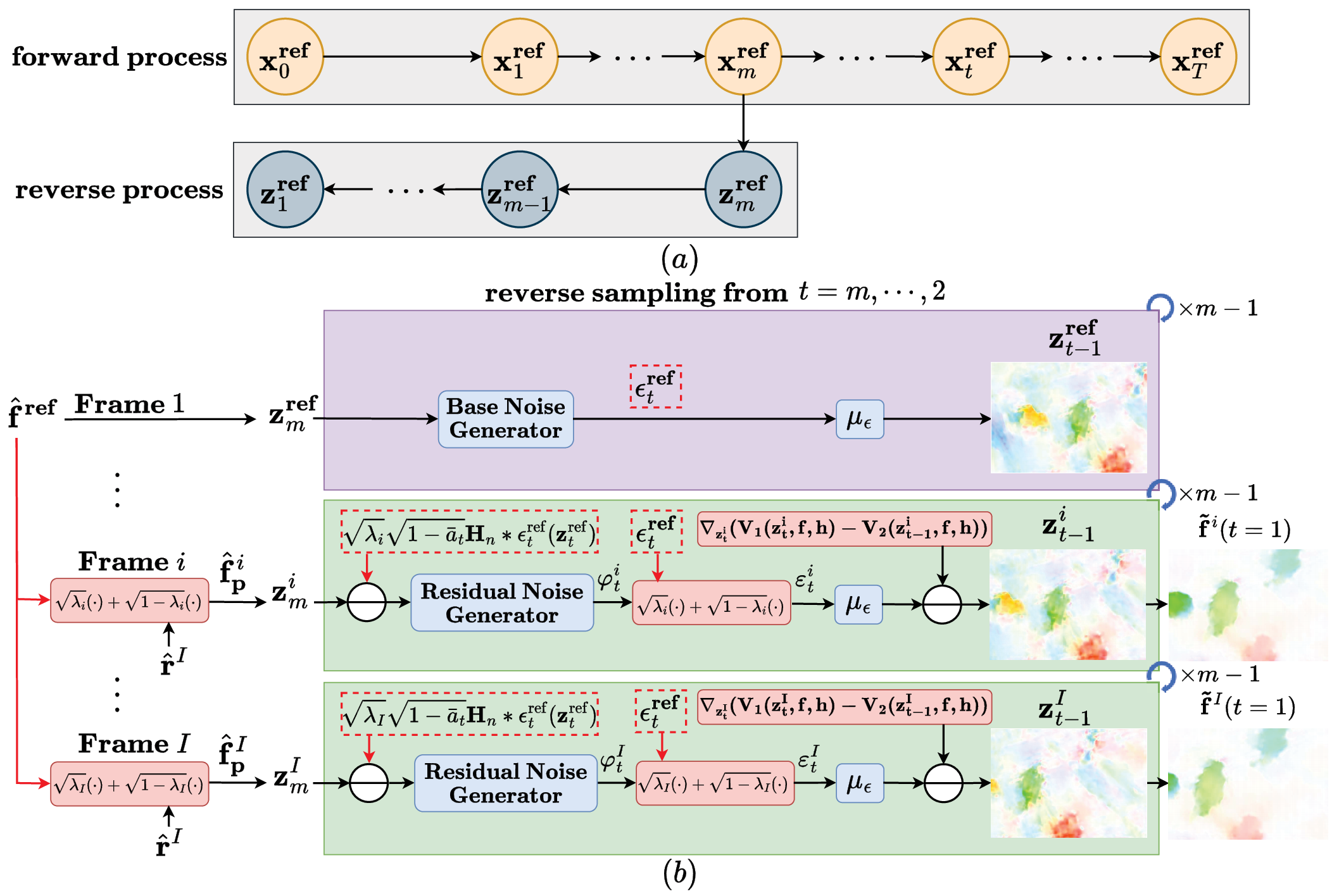}
	\caption{(a) The diffusion process for the semantic I frame. (b) The proposed DDMFC module. The semantic I frame shares both the base frame and base noise to other semantic P frames in a GoP. While other semantic P frames produce unique residual noise and multi-frame condition steering to generate corresponding semantic frames at the receiver. $\mu_{\epsilon}$ denotes mean-value predicted function of DDIM in $\epsilon$-prediction formulation.}
	\label{fig_4}
\end{figure*}

Although the EBM-based GMFC strategy facilitates the conditional generation of current semantic frames, it encounters a significant challenge in maintaining temporal consistency across frames within the same GoP, which imposes a strict constraint on effective frame compensation. This is primarily because an estimated semantic P frame shares substantial similarity with its received semantic I frame, necessitating the suppression of generative diversity to avoid visual incoherence and artifacts. Furthermore, the GMFC approach does not account for the residuals of each semantic frame during the compensation process, overlooking valuable high-frequency details that could enhance reconstruction fidelity. To holistically address these limitations by simultaneously enforcing robust temporal consistency as inspired by \cite{videofusion} while incorporating residual-aware refinement, we propose a decoupled diffusion multi-frame compensation module. This advanced solution builds upon the conditional steering of GMFC by employing a dedicated diffusion process to expertly manage the frame compensation task, thereby ensuring both the structural integrity and detailed accuracy of the generated video sequences. We first decouple the diffusion target of semantic P frames into both reference semantic frame and corresponding residuals, which is illustrated as
\begin{align}
	\mathbf{f}^i_{\mathrm{p}} = \sqrt{\lambda_i}\mathbf{f}^\mathrm{ref}+\sqrt{1-\lambda_i}\mathbf{r}^i,    
\end{align}
where $\lambda_i \in [0,1]$ denotes the weight parameter which controls the generation difference between the semantic I frame and the current semantic P frame. 

Analogous to Eq. (12), the estimated noise $\mathbf{\varepsilon}_t^i(\mathbf{z'}_t^i,\mathbf{z}_{t}^\mathrm{ref})$ can also be decomposed into two concatenated learnable noises as the diffusion-based methods do \cite{videofusion}, which is represented as
\begin{align}
\mathbf{\varepsilon}_t^i(\mathbf{z'}_t^i,\mathbf{z}_{t}^\mathrm{ref})=\sqrt{\lambda_i}\mathbf{\epsilon}_t^\mathrm{ref}(\mathbf{z}_{t}^\mathrm{ref})+\sqrt{1-\lambda_i}\mathbf{\varphi}_t^i(\mathbf{z'}_t^i),
\end{align}
where $\mathbf{\varepsilon}_t^i(\mathbf{z'}_t^i,\mathbf{z}_{t}^\mathrm{ref})$ and $\mathbf{\varphi}_t^i(\mathbf{z'}_t^i)$ refer to the total noise and residual noise for the current semantic P frame, respectively. $\mathbf{z'}_t^i$ refers to the polished sample.

With the decoupled frame and noise, we then provide the diffusion process for generating P frames with only reference I frame and corresponding residuals. Following the denoised diffusion forward process, since the actual P frame is unknown at the receiver, the forward start point $\mathbf{z}_s^i$ can be given as
\begin{align}
	\mathbf{z}_s^i=\mathbf{H}_s\ast\mathbf{f}^i_\mathrm{p}.
\end{align}
where $\mathbf{z}_s^i$ encapsulates the rough P frame composed of base frame $\mathbf{f}^\mathrm{ref}$ and current temporal information $\mathbf{r}^i$ along with the channel equalization matrix $\mathbf{H}_s$. 

With Eq. (12) and Eq. (13), the diffusion forward process can be rewritten as
\begin{equation}
	\begin{aligned}
		\mathbf{z}_{t}^{i}&=\sqrt{\bar{a}_t}\mathbf{z}_s^i+\sqrt{1-\bar{a}_t}\mathbf{H}_n\ast\varepsilon_t^i(\mathbf{z}_t^i,\mathbf{z}_{t}^\mathrm{ref}) \\& =\sqrt{\bar{a}_t}\mathbf{H}_s\ast(\sqrt{\lambda_i}\mathbf{f}^\mathrm{ref}+\sqrt{1-\lambda_i}\mathbf{r}^i)\\&+\sqrt{1-\bar{a}_t}\mathbf{H}_n\ast(\sqrt{\lambda_i}\mathbf{\epsilon}_t^\mathrm{ref}(\mathbf{z}_{t}^\mathrm{ref})+\sqrt{1-\lambda_i}\mathbf{\varphi}_t^i(\mathbf{z'}_t^i)) \\& =\sqrt{\lambda_i}\underbrace{(\sqrt{\bar{a}_t}\mathbf{H}_s\ast\mathbf{f}^\mathrm{ref}+\sqrt{1-\bar{a}_t}\mathbf{H}_n\ast\mathbf{\epsilon}_t^\mathrm{ref}(\mathbf{z}_{t}^\mathrm{ref}))}_{\text{diffusion of $\mathbf{f}^\mathrm{ref}$}}\\&+\sqrt{1-\lambda_i}\underbrace{(\sqrt{\bar{a}_t}\mathbf{H}_s\ast\mathbf{r}^i+\sqrt{1-\bar{a}_t}\mathbf{H}_n\ast\mathbf{\varphi}_t^i(\mathbf{z'}_t^i))}_{\text{diffusion of $\mathbf{r}^i$}}.
	\end{aligned}
\end{equation}

It is clearly that the whole diffusion process is split into two parts. One is the reference semantic frame generation. The other is the residual generation of the current semantic frame. Follow this way, the diffusion result of $\mathbf{f}^\mathrm{ref}$ can be shared among other frames in a GoP. Moreover, $\mathbf{\epsilon}_t^\mathrm{ref}(\mathbf{z}_{t}^\mathrm{ref})$ is assumed as the base noise for the generation process of other frames in the same GoP.

For every reverse sampling step, the base noise from semantic I frame is first learnt. As such, for the residual noise computation, the effect of base noise is first removed as
\begin{align}
	\mathbf{z'}_{t}^{i}=\mathbf{z}_{t}^{i}-\sqrt{\lambda_i}\sqrt{1-\bar{a}_t}\mathbf{H}_n\ast\mathbf{\epsilon}_t^\mathrm{ref}(\mathbf{z}_{t}^\mathrm{ref}),i\ne1
\end{align}

In this way, the estimated diffusion noise is formulated as
\begin{align}
\mathbf{\varepsilon}_t^i(\mathbf{z'}_t^i,\mathbf{z}_{t}^\mathrm{ref})=\begin{cases}
	\mathbf{\epsilon}_t^\mathrm{ref}(\mathbf{z}_t^{\mathrm{ref}}), i=1\\
	\sqrt{\lambda_i}\mathbf{\epsilon}_t^\mathrm{ref}(\mathbf{z}_t^{\mathrm{ref}})+\sqrt{1-\lambda_i}\mathbf{\varphi}_t^i(\mathbf{z'}_t^i), i\ne1
\end{cases}
\end{align}

Finally, analogous to Eq. (11), the reverse sampling for the decoupled diffusion is described as
\begin{equation}
	\begin{aligned}
		\mathbf{z}_{t-1}^i &= \sqrt{\bar{a}_{t-1}}( \frac{\mathbf{z}_{t}^i-\sqrt{1-\bar{a}_t}\mathbf{H}_n\ast\mathbf{\varepsilon}_t^i(\mathbf{z'}_t^i,\mathbf{z}_{t}^\mathrm{ref})}{\sqrt{\bar{a}_t}}) \\&+\sqrt{1-\bar{a}_{t-1}-\sigma^2_t} \mathbf{H}_n\ast\mathbf{\varepsilon}_t^i(\mathbf{z'}_t^i,\mathbf{z}_{t}^\mathrm{ref})+\sigma_t\mathbf{\epsilon} \\&-k(t)\nabla _{\mathbf{z}_t^i}(V_1(\mathbf{z}_t^i,f,\mathbf{h})-V_2(\mathbf{z}_{t-1}^i,f,\mathbf{h})).  
	\end{aligned}
\end{equation}

Furthermore, to select the proper start point $\mathbf{z}_m^i$ for Eq. (18), we first derive the distribution of diffusion forward process from Eq. (15) as
\begin{align}
	q(\mathbf{z}_m^i|\mathbf{z}_s^i,\mathbf{h})\sim \mathcal{N}(\mathbf{z}_m^i;\sqrt{\bar{a}_m}\mathbf{z}_s^i,(1-\bar{a}_m)\mathbf{H}^2_n),
\end{align}
where $\mathbf{z}_m^i$ refers to the diffusion sample at the $m$-th timestep.

Then the reconstructed forward start point $\mathbf{\hat{f}}^i_\mathrm{p}$ is given as
\begin{equation}
	\begin{aligned}
	\mathbf{\hat{f}}^i_\mathrm{p}&=\sqrt{\lambda_i}\mathbf{\hat{f}}^\mathrm{ref}+\sqrt{1-\lambda_i}\mathbf{\hat{r}}^i\\&=\sqrt{\lambda_i}(\mathbf{H}_s\ast\mathbf{f}^{\mathrm{ref}}+\mathbf{H}_n\ast\mathbf{n})+\sqrt{1-\lambda_i}(\mathbf{H}_{s}\ast\mathbf{r}^i+\mathbf{H}_{n}\ast\mathbf{n}_\mathbf{r}^i)\\&=\mathbf{H}_s(\underbrace{\sqrt{\lambda_i}\mathbf{f}^{\mathrm{ref}}+\sqrt{1-\lambda_i}\mathbf{r}^i}_{\mathbf{f}^i_{\mathrm{p}}})+\mathbf{H}_n(\underbrace{\sqrt{\lambda_i}\mathbf{n}+\sqrt{1-\lambda_i}\mathbf{n}_\mathbf{r}^i}_{\mathcal{N}(0,\sigma^2)})\\&=\mathbf{H}_s\mathbf{f}^i_\mathrm{p}+\mathbf{H}_n\delta^i.
	\end{aligned}
\end{equation}
where $\delta^i\sim\mathcal{N}(0,\sigma^2)$ refers to the Gaussian noise for the $i$-th semantic P frame.

To this end, the distribution of normalized $\mathbf{\hat{f}}^i_\mathrm{p}$ is provided as
\begin{align}
	p(\mathbf{\tilde{f}}^i_\mathrm{p}|\mathbf{z}_s^i,\mathbf{h})\sim \mathcal{N}(\mathbf{\tilde{f}}^i_\mathrm{p};\frac{1}{\sqrt{1+\sigma^2}}\mathbf{z}_s^i,\frac{\sigma^2}{1+\sigma^2}\mathbf{H}^2_n),
\end{align}
where $\mathbf{\tilde{f}}^i_\mathrm{p}=\frac{1}{\sqrt{1+\sigma^2}}\mathbf{\hat{f}}^i_\mathrm{p}$ is the normalized $\mathbf{f}^i_\mathrm{p}$ frame.

Based on Eq. (19) and Eq. (21), it is easy to observe that if $\bar{a}_m=\frac{1}{1+\sigma^2}$, the KL-Divergence turns to be
\begin{align}
	D_{\mathrm{KL}}(q(\mathbf{z}_m^i|\mathbf{z}_s^i,\mathbf{h})||p(\mathbf{\tilde{f}}^i_\mathrm{p}|\mathbf{z}_s^i,\mathbf{h}))=0.
\end{align}

In this way, rather than start from a random Gaussian noise, $\mathbf{\tilde{f}}^i_\mathrm{p}$ can be assumed as the start point for the reverse sampling with $m$ steps towards $\mathbf{z}_s^i$. However, it is not obliged to take full sampling steps since $\mathbf{f}^i_{\mathrm{p}}$ is simply an approximation of current semantic P frame. To this end, only very small sampling steps enable the frame generation towards current semantic P frames through multi-frame conditional steering and decoupled diffusion process.

For the final diffusion reverse sampling step $t=1$, the generated current semantic P frame $\tilde{\mathbf{f}}^i$ is given as
\begin{align}
\tilde{\mathbf{f}}^i=\frac{\mathbf{z}_{1}^i-\sqrt{1-\bar{a}_1}\mathbf{H}_n\ast\mathbf{\varepsilon}_1^i(\mathbf{z'}_1^i,\mathbf{z}_{1}^\mathrm{ref})}{\sqrt{\bar{a}_1}}
\end{align}

The whole process of proposed DDMFC module is summarized in Fig. \ref{fig_4}(b). For an arbitrary GoP, the received semantic I frame is assumed as the base frame for producing the start point for reverse sampling. For every timestep $t$, the predicted noise $\mathbf{\epsilon}_t^\mathrm{ref}$ from the semantic I frame is shared across other semantic P frames for generating $\mathbf{\varepsilon}_t^i(\mathbf{z'}_t^i,\mathbf{z}_{t}^\mathrm{ref})$. The uniqueness of each semantic P frame is retained through different diffusion start points, produced residual noise, and condition steering. With the iteration of $m$ steps, the negative effects of channel impairment and lack of temporal information among adjacent frames can be mitigated a lot through generating successive semantic frame samples for each current state.

With the above designs, WVSC-D framework which combines semantic communication with the deep video coding structure enables efficient wireless video transmission.

\section{Deployment of WVSC-D Framework}
In this section, we provide the deployment details of WVSC-D framework, including network structure, training loss, and training strategy, respectively.

\begin{figure}[htbp]
	\centering
	\includegraphics[width=3.5in]{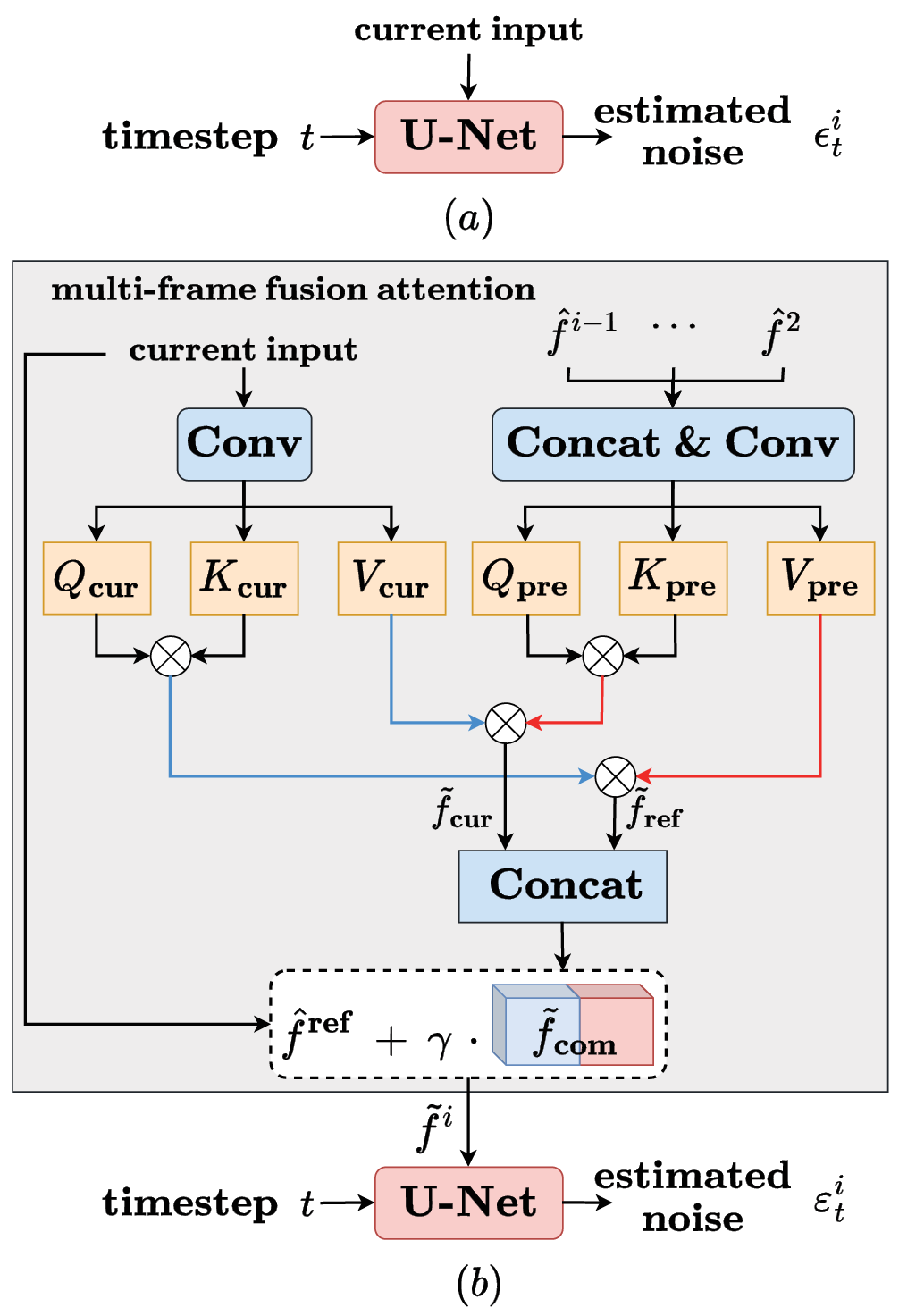}
	\caption{The architecture of diffusion network. (a) For the semantic I frame: U-Net (base noise generator). (b) For the semantic P frame: U-Net with MFA module (residual noise generator).}
	\label{fig_5}
\end{figure}

\subsection{Network Structure}
For the semantic codec $f_e(\cdot)$ and $g_a(\cdot)$, the Swin-Transformer \cite{swin} is adapted as the network backbone.

For the decoupled diffusion multi-frame compensation module, the U-net \cite{unet} structure is adapted to generate the estimated base noise for every timestep $t$, as shown in Fig. \ref{fig_5}(a) for the semantic I frame. For the semantic P frame, to fuse the previously reconstructed frames as conditions into the diffusion process, a MFA module is designed to produce the conditional input for the diffusion network.

The detailed architecture of proposed diffusion network is shown in Fig. \ref{fig_5}(b). The structure divides the computation process into two steps. The initial step is the multi-frame attention fusion. Two pairs of attention matrices $\mathbf{Q}_{\mathrm{cur}}$, $\mathbf{K}_{\mathrm{cur}}$, $\mathbf{V}_{\mathrm{cur}}$ and $\mathbf{Q}_{\mathrm{pre}}$, $\mathbf{K}_{\mathrm{pre}}$, $\mathbf{V}_{\mathrm{pre}}$ are generated based on the current input and several previous semantic frames $\mathbf{\hat{f}}$, respectively. Then through multiplication between cross attention matrices \cite{cross}, the previous semantics $\mathbf{\tilde{f}}_{\mathrm{pre}}$ and current input are computed. Finally, through residual connection, the compensated current semantic frame is presented. The whole process can be formulated as
\begin{align}
	\mathbf{\tilde{f}}_{\mathrm{cur}} & = \Phi(\mathbf{Q}_{\mathrm{cur}}\mathbf{K}_{\mathrm{cur}}^T)\mathbf{V}_{\mathrm{pre}},
\end{align}
\begin{align}
	\mathbf{\tilde{f}}_{\mathrm{pre}} & = \Phi(\mathbf{Q}_{\mathrm{pre}}\mathbf{K}_{\mathrm{pre}}^T)\mathbf{V}_{\mathrm{cur}},
\end{align}
\begin{align}
	\mathbf{\tilde{f}}_{\mathrm{com}} & = \mathrm{Con}(\mathbf{\tilde{f}}_{\mathrm{pre}}, \mathbf{\tilde{f}}_{\mathrm{cur}}),
\end{align}
\begin{align}
	\mathbf{\tilde{f}}^i & = \mathbf{\hat{f}}^{\mathrm{ref}} + \gamma \mathbf{\tilde{f}}_{\mathrm{com}},
\end{align}
where $\Phi(\cdot)$ is the softmax function, $\mathrm{Con}(\cdot,\cdot)$ refers to the concatenation between two feature matrices, $\gamma$ denotes the weight parameter.

\subsection{Training Loss}
The training loss of the WVSC-D contains two parts. One is the end-to-end video reconstruction loss, where mean square error (MSE) is set as the default loss.
\begin{align}
	L_\mathrm{R}=\frac{1}{N\times I}\sum_{n=1}^{N}\sum_{i=1}^{I}||\hat{\mathbf{x}}_n^i-{\mathbf{x}}_n^i||^2,
\end{align}
where $\hat{\mathbf{x}}_n^i$ and ${\mathbf{x}}_n^i$ refer to the $i$-th reconstructed video frame and original frame in the $n$-th GoP, respectively.

The other is the diffusion training loss. According to Eq. (18), the training loss for the DDMFC module for an arbitrary semantic frame is given as
\begin{align}
	L_{n_t}^i=\begin{cases}||\mathbf{\epsilon}-\mathbf{\epsilon}_{n_t}^i(\mathbf{z}_{n_t}^{\mathrm{ref}})||^2, i=1
		\\||\mathbf{\epsilon}-\sqrt{\lambda_i}[\mathbf{\epsilon}_{n_t}^i(\mathbf{z}_{n_t}^{\mathrm{ref}})]_{sg}-\sqrt{1-\lambda_i}\mathbf{\varphi}_{n_t}^i(\mathbf{z'}_{n_t}^i)||^2, i\ne1
	\end{cases}
\end{align}
where $L_{n_t}^i$ refers to the diffusion training loss of the $i$-th frame in the $n$-th GoP for the random timestep $t$. $\mathbf{\epsilon}_{n_t}^i(\mathbf{z}_{n_t}^{\mathrm{ref}})$ and $\mathbf{\varphi}_{n_t}^i(\mathbf{z'}_{n_t}^i)$ refer to the corresponding base noise and residual noise, respectively. $[\cdot]_{sg}$ refers to the stop gradient, which means that the diffusion training for the current frame does not abrupt the corresponding reference semantic frame.

Thus, the total training loss for the DDMFC is 
\begin{align}
L_\mathrm{D}=\frac{1}{N\times I}\sum_{n_t=1}^{N}\sum_{i=1}^{I}L_{n_t}^i.
\end{align}

Finally, the loss of WVSC-D is collected as
\begin{align}
	L_\mathrm{W}=L_\mathrm{R}+\mu L_\mathrm{D}.
\end{align}
where $\mu$ refers to the trade-off term with a small value for the JSCC framework and the DDMFC.

\begin{algorithm}[htbp]
	\caption{Sampling algorithm of DDMFC}\label{alg:alg2}
	\begin{algorithmic}
		\STATE 
		$\textbf{Input:}$ received reference semantic frame $\mathbf{\hat{f}}^\mathrm{ref}$, received residuals $\mathbf{\hat{r}}$, previous multiple received frames $\mathbf{f}$, channel response $\mathbf{h}$
		\STATE
		$\textbf{Output:}$ Compensated current semantic P frame $\tilde{\mathbf{f}}$
		
		
		\STATE \hspace{0.5cm}$ \textbf{} $
		
		\STATE1. $\mathbf{z}_{m}^\mathrm{ref}=\mathbf{\hat{f}}^\mathrm{ref}, i=1$
		\STATE2. $\mathbf{z}_{m}^i=\mathbf{\tilde{f}}^i_\mathrm{p}, i=2,\cdots,N$ 
		\STATE3. \textbf{for $t=m,\cdots,1$ do}
		\STATE4. \hspace{0.4cm} \textbf{if $i = 1$} \text{(for the base noise generation)}
		\STATE5. \hspace{0.8cm} \text{Compute $\mathbf{z}_{t-1}^\mathrm{ref}$ according to Eq. (6)}
		\STATE6. \hspace{0.8cm} \text{Acquire each base noise $\mathbf{\epsilon}_t^\mathrm{ref}$}		
		\STATE7. \hspace{0.4cm} \textbf{else} \text{(for the semantic P frame generation)}
		\STATE8. \hspace{0.8cm} \textbf{if $t = 1$}
		\STATE9. \hspace{1.2cm} $\tilde{\mathbf{f}}^i=\frac{\mathbf{z}_{1}^i-\sqrt{1-\bar{a}_1}\mathbf{H}_n\ast\mathbf{\varepsilon}_1^i(\mathbf{z'}_1^i,\mathbf{z}_{1}^\mathrm{ref})}{\sqrt{\bar{a}_1}}$
		\STATE10. \hspace{0.6cm} \textbf{else}
		\STATE11. \hspace{1.0cm} \text{Compute $\mathbf{z}_{t-1}^i$ according to Eq. (18)}
		\STATE12. \textbf{end for}
		
	\end{algorithmic}
	\label{alg2}
\end{algorithm}

\subsection{Training and Sampling Strategy}
Since the JSCC part and the DDMFC module are two framework components with different training targets, it is hard to acquire a well-trained framework through a single-stage training method. To exploit the advantages of such semantic communication framework with diffusion-based multi-frame compensation, we adapt a three-stage training strategy. In the first stage, since the P frame reconstruction requires the multi-frame compensation, to achieve a well-trained JSCC encoder, the JSCC network is joint trained with the compensation module. The loss function is chosen as $L_\mathrm{W}$. In the second stage, we fix the JSCC network and only train the compensation module. The loss function is $L_\mathrm{D}$. In the last stage, with the pretrained compensation module, we assume it as a plug-in module for the whole network and solely train the JSCC decoder while set the JSCC encoder and the compensation module fixed. The loss function is $L_\mathrm{R}$.

For the DDMFC module, the training method is the same as in \cite{ddpm} while the diffusion sampling method is provided in Alg. 1. The $\mathbf{z}_{m}^\mathrm{ref}$ is provided according to \cite{cddm}. Note that for the wireless video transmission, we set the conditional steering function $V_1(\mathbf{z}_t^i,f,\mathbf{h})$ as MSE and $V_2(\mathbf{z}_t^{i-1},f,\mathbf{h})$ as 0, respectively.

\section{Numerical Results}
In this section, we present numerical results to evaluate the effectiveness of proposed WVSC-D for wireless video transmission.

\subsection{Experimental Setups}

\subsubsection{Datasets}
For the wireless video semantic transmission, we quantify the performances of proposed WVSC-D versus other benchmarks over the UCF101 \cite{ucf} dataset, which is a widely-used video dataset. Since it has longer duration and hundreds of frames in each video clip, it is especially suitable for evaluating the temporal consistency for reconstructed videos. The dataset is split into about 5:1 ratio for training and testing, respectively. During model training, images are randomly cropped to 128$\times$128$\times$3.

\subsubsection{Model Deployment Details}
The network deployment of WVSC-D utilizes the WITT \cite{witt} backbone as the semantic codec with $\{N_1,N_2,N_3,N_4\}=\{2,2,2,2\}$ Swin-Transformer blocks \cite{swin}. WITT is also adopted as the I frame reconstruction network and plugged into WVSC-D with pretrained weights. Various signal-to-noise ratios (SNRs) are randomly sampled for each image batch during training stage to acquire SNR-adaptive results in various test SNR conditions. For model training, we use variable learning rate, which decreases step-by-step from 1e-4 to 2e-5 along with different epochs. The batchsize is set as 1. The training GoP size $N$ is set as 10. The trade-off term $\mu$ is set as 1e-4. The whole framework is optimized with AdamW \cite{adamw} algorithm. Under the Rayleigh fading channel conditions, DL-based schemes employ the minimum mean square error (MMSE) \cite{mmse} algorithm for channel equalization. The total timestep $T$ is set as 1000 and the default sampling step $m$ is set as 10. To provide a satisfying balance between continuity and uniqueness for DDMFC, every $\lambda_i$ is set as 0.7 and $k(t)$ is set as 0.3. To ensure the stability of diffusion generation, the noise variance $\sigma_t$ is set as 0. All the experiments are run in RTX 3090 GPUs with Pytorch2.0.0.

\subsubsection{Comparison Benchmarks}
In the experiments, several benchmarks are given as below:

$\textbf{DVSC}$: The DL-empowered deep video transmission framework \cite{dvsc} with SNR-adaptive channel coder and semantic repairment at the receiving end.

$\textbf{WVSC-G}$: The wireless video semantic transmission framework which only utilizes proposed GMFC strategy in Section III.B for multi-frame compensation.

$\textbf{WVSC}$: The semantic-level wireless video semantic transmission framework \cite{wvsc} which directly utilizes MFA module in Section IV.A for multi-frame compensation.


$\textbf{H.264+LDPC+QAM}$: The SSCC scheme with standardized H.264 \cite{264} video codec as the source coding scheme and 5G LDPC as the channel coding scheme, along with the quadrature amplitude modulation (QAM).

$\textbf{H.265+LDPC+QAM}$: The SSCC scheme with standardized H.265 \cite{265} video codec as the source coding scheme and 5G LDPC as the channel coding scheme, along with the QAM.

Note that DVSC is the DL-based pixel-level point-to-point wireless transmission scheme, which corresponds to Fig. 1(b) which directly conduct video coding in signal level. Traditional H.265/H.264+LDPC+QAM are SSCC schemes which H.265 and H.264 are employed by ffmpeg-python \cite{ffmpeg} with low delay pattern of GoP size 32 while 5G LDPC with code length 4096 along with QAM is implemented by sionna \cite{sionna}. The code rate of 5G LDPC and modulation order of QAM are varied according to different SSCC types. WVSC-G and WVSC are ablation benchmarks compared to WVSC-D.

\subsubsection{Evaluation Metrics}

We leverage the widely used pixel-wise metric peak signal-to-noise ratio (PSNR), perceptual-level multi-scale structural similarity (MS-SSIM) \cite{ssim} and learned perceptual image patch similarity (LPIPS) \cite{lpips} as measurements for the reconstructed image quality. 

\begin{figure*}[htbp]
	\centering  
	\subfigure[PNSR for the reconstructed videos.]{
		\includegraphics[width=0.32\linewidth]{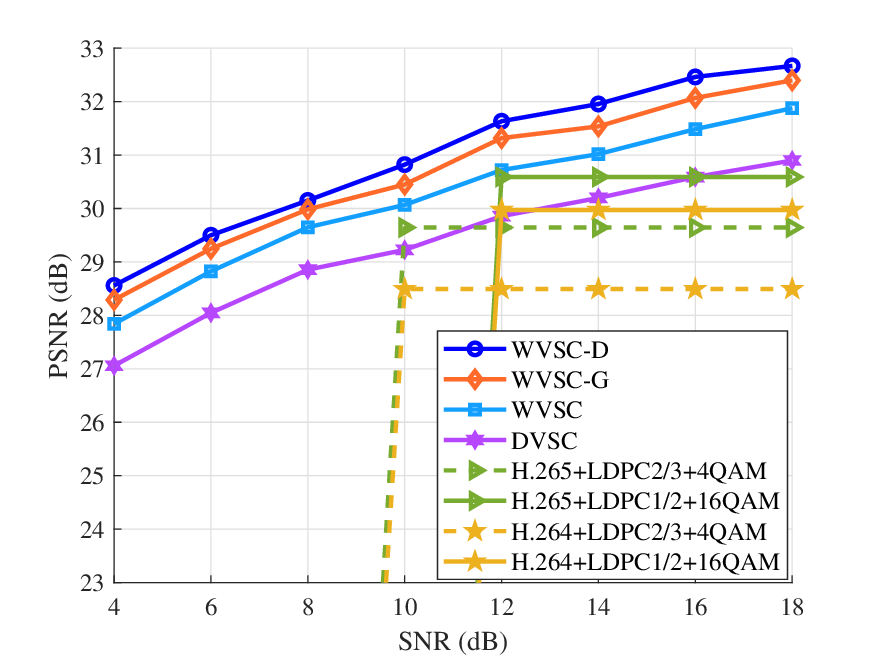}}
	\subfigure[MS-SSIM for the reconstructed videos.]{
		\includegraphics[width=0.32\linewidth]{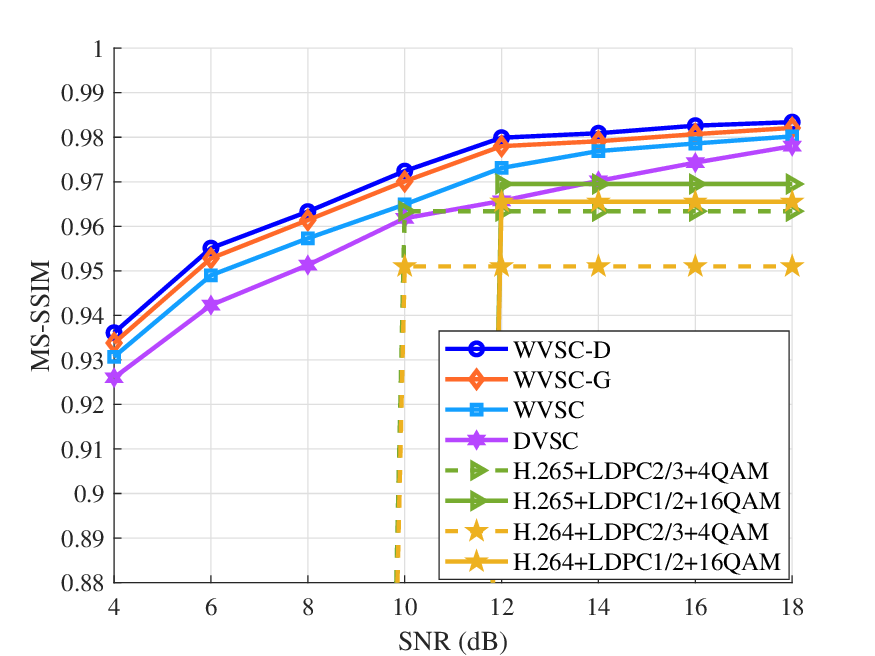}}
	\subfigure[LPIPS for the reconstructed videos.]{
		\includegraphics[width=0.32\linewidth]{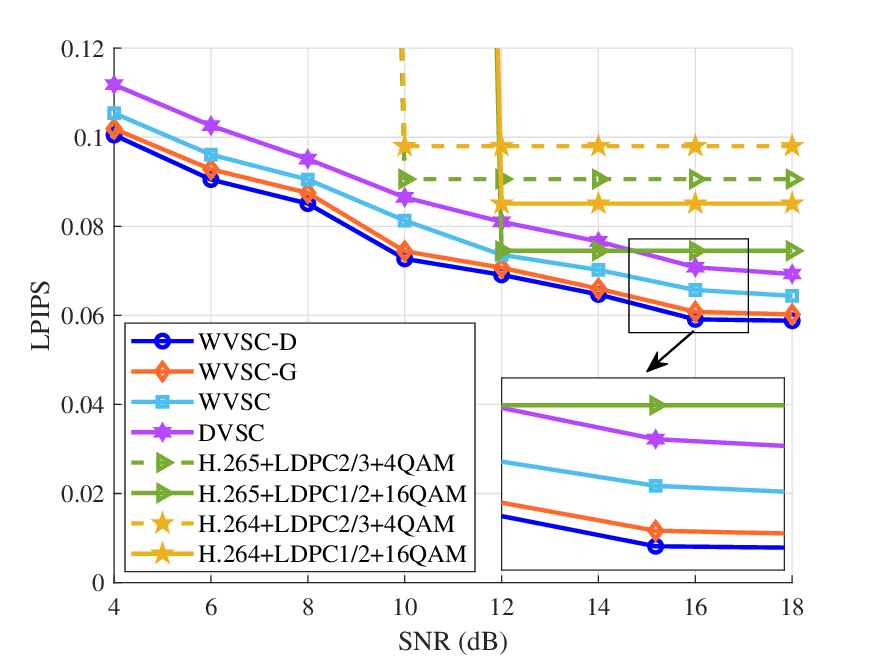}}
	\caption{Quality of the reconstructed images versus the SNRs under Rayleigh fading channels (CBR = 0.04).}
	\label{fig_6}
\end{figure*}
\begin{figure*}[htbp]
	\centering  
	\subfigure[PNSR for the reconstructed videos.]{
		\includegraphics[width=0.32\linewidth]{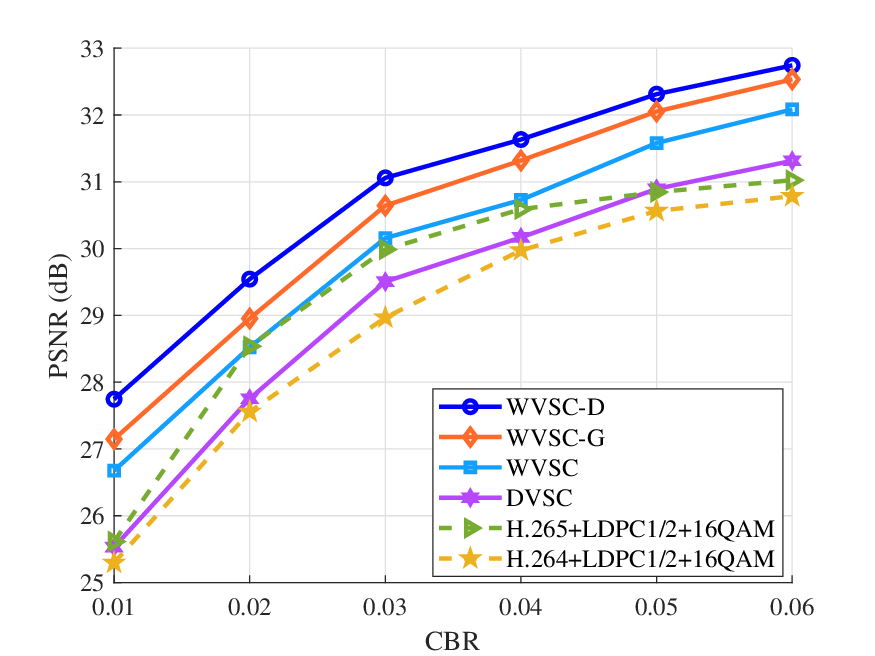}}
	\subfigure[MS-SSIM for the reconstructed videos.]{
		\includegraphics[width=0.32\linewidth]{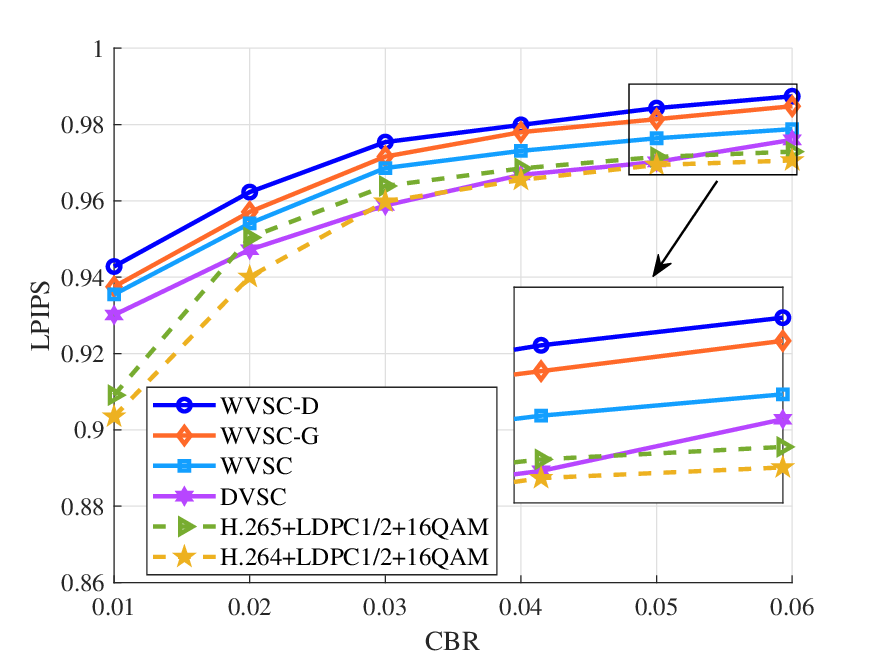}}
	\subfigure[LPIPS for the reconstructed videos.]{
		\includegraphics[width=0.32\linewidth]{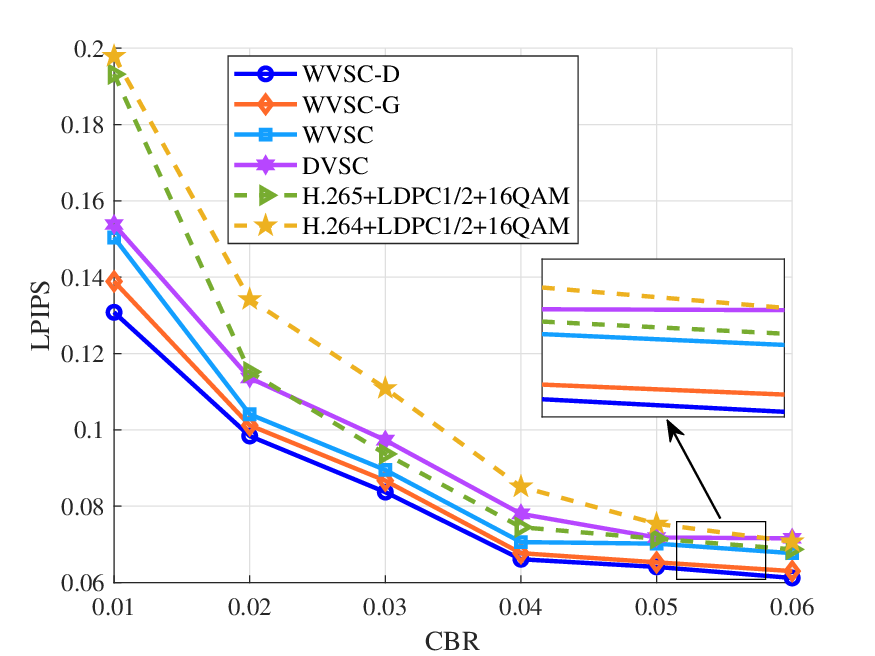}}
	\caption{Quality of the reconstructed images versus the CBRs under Rayleigh fading channels (SNR = 12 dB).}
	\label{fig_7}
\end{figure*}

\subsection{Results Analysis}

\subsubsection{Performance for Different SNRs}
We first evaluate the anti-noise performances of WVSC-D under Rayleigh fading channels with a specific channel bandwidth ratio (CBR). As shown in Fig. \ref{fig_6}(a), it is clearly to observe that WVSC-D outperforms all other benchmarks. Compared to the pixel-level transmission scheme, WVSC-D outperforms DVSC for about 1.8 dB in terms of PSNR. This trend verifies that semantic-level video coding-based deep video delivery structure seems to be more robust to the channel fading and noise. For the traditional separated coding schemes, two channel coding and modulation parameter pairs are presented. In detail, 2/3 code rate LDPC with 4QAM and 1/2 code rate LDPC with 16QAM are set, which reflect corresponding anti-noise performances under different channel interference levels. Compared to traditional schemes, WVSC-D provides satisfying performance gain and stability since traditional schemes would be confronted with serious cliff effect with harsh channel conditions. As shown in Fig. \ref{fig_6}(b) and Fig. \ref{fig_6}(c), the DL-based schemes in general achieves better reconstruction results in terms of MS-SSIM and LPIPS compared to traditional separated coding schemes, which means that the satisfying visual perception quality is ensured. Compared to DVSC, WVSC-D preserves even much more high frequency image details. These results demonstrate the effectiveness of WVSC-D for video delivery under Rayleigh fading channels with various noise intensities.

For the ablation benchmarks, WVSC-D also outperforms WVSC-G and WVSC. For the WVSC, it employs cross attention-based multi-frame compensation to fuse previously reconstructed frames as extra temporal information into the current frame compensation. However, such feature extraction-based network design presents insufficient ability for generating current semantic P frames at the receiver. For the WVSC-G, GMFC module is adapted to generate the polished semantic frame through conditional diffusion reverse sampling process. The strong generation ability of diffusion along with the conditional steering helps polish the semantic I frame into current semantic P frame. Generally about 0.8 dB over the WVSC. DDMFC, which decouples the frame and diffusion process to ensure the temporal consistency during frame generation, achieves further performance gain compared to the GMFC diffusion strategy.

\subsubsection{Performance for Different CBRs}
Then we evaluate the bandwidth compression performances of WVSC-D under Rayleigh fading channels with SNR = 12 dB. As shown in Fig. \ref{fig_7}(a), WVSC-D achieves much performance gain compared to other schemes. Compared to DVSC, the performance gap increases as the increase of CBRs. Such insight mainly lies in the deployment of video coding in semantic level. Since semantic coding has great potential of source compression, based on the extracted semantics, the video coding enables much more effective data compression performances compared to directly conduct on video signals. In this way, higher CBRs allow the video coding part to retain more semantic information in the form of frame residuals. As the reference frames stay relatively steady as CBR values change, the increased allocation of extra bandwidth resources for transmitting residuals greatly promotes the reconstruction of each semantic frame. For the visual perceptual-level indexes in Fig. \ref{fig_7}(b) and Fig. \ref{fig_7}(c), WVSC-D retains relatively satisfying visual quality for a wide range of CBRs compared to other schemes. 

\subsubsection{Visualization Results for the Wireless Video Transmission}
\begin{figure*}[htbp]
	\centering
	\includegraphics[width=6.8in]{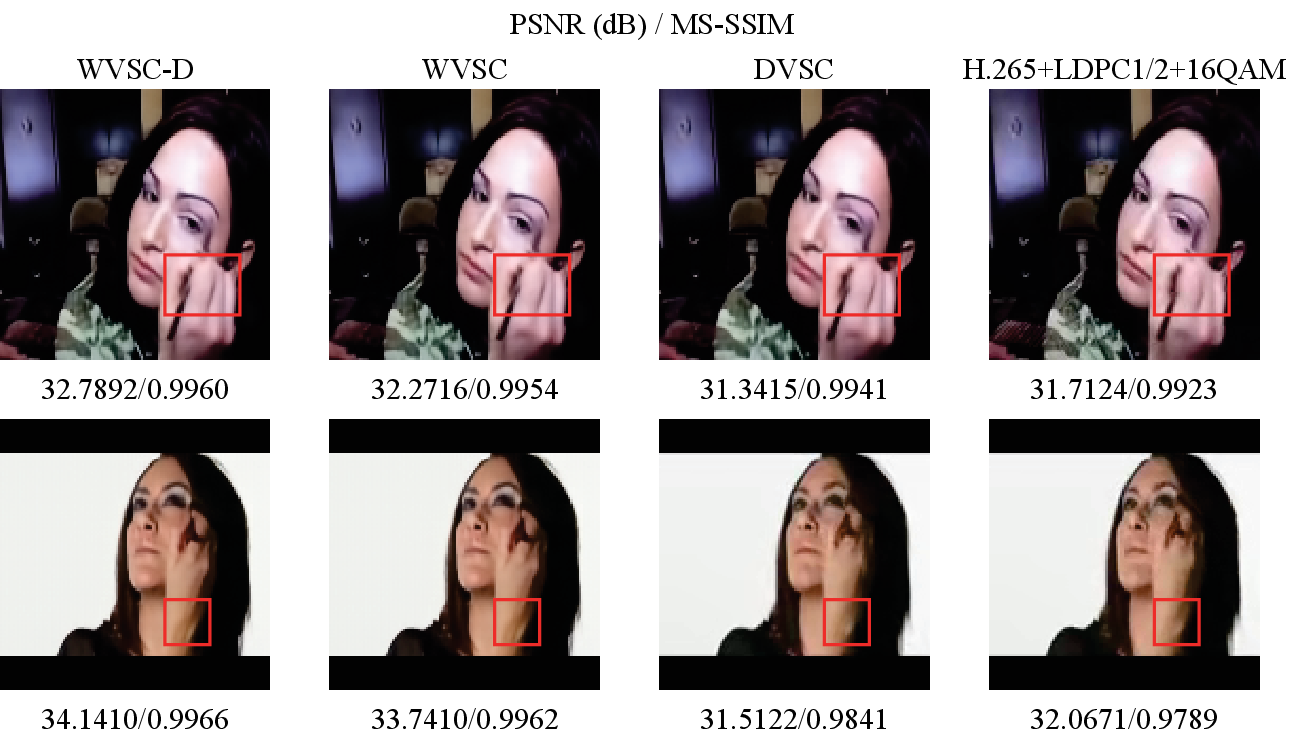}
	\caption{Visualized results for the WVSC-D and other benchmarks. (SNR = 12 dB, CBR = 0.10)}
	\label{fig_8}
\end{figure*}

We present the visualization results for the wireless video transmission in Fig \ref{fig_8}. In general, DL-based schemes outperform the traditional schemes for not only evaluation indexes such as PSNR and MS-SSIM but also visual quality since neural networks can retain more high frequency video signal details. It is evidently to observe that traditional schemes are unable to avoid blurry points for reconstructed videos while DL-based schemes provide more smooth images for different video clips. For other DL-based schemes such as WVSC and DVSC, WVSC-D achieves better performances. With proposed WVSC-D, reconstructed videos with sound visual reconstructed quality are provided. 

\subsubsection{Analysis for the Hyperparameters in the DDMFC Module}
We next present a sensitive analysis for the hyperparameters in the DDMFC module, which are set manually. As shown in Tab. \ref{table2}, we evaluate the impact of varying the sampling step $m$. Results indicate that larger values of $m$ yield better reconstruction performance at the cost of increased computational load. It is therefore crucial to balance these two factors. A default value of $m=10$ offers a strong performance gain over $m=5$, while the marginal improvement from $m=10$ to $m=15$ is significantly smaller. This trend confirms that $m=10$ represents an effective trade-off for the diffusion reverse sampling process.

Then, we analyse the weight parameter $\lambda_i$, which controls the influence of the semantic I frame on the generated P frame. A high $\lambda_i$ value for frame compensation enforces strong similarity to the semantic I frame, resulting in generated frames with slight motion. Conversely, a low $\lambda_i$ value allows for greater diversity but can lead to unstable and inaccurate semantic P frame reconstruction. Since the goal is to transmit a coherent video clip where frames within a GoP are highly similar, we set a default value of $\lambda_i=0.7$. As shown in Tab. \ref{table3}, small values ($\lambda_i=0.01, 0.3$) introduce excessive variation and instability. A very high value ($\lambda_i=0.99$) enforces near-complete consistency with the semantic I frame and sacrifices crucial motion details.

\begin{table}[htbp]
	\centering
	\caption{Sensitive analysis of sampling step $m$.}
	\label{table2}
	
	\begin{tabular}{|c|c|c|c|c|}  
		\hline 
		& & & &\\[-6pt] 
		$m$&1&5&10&15 \\
		\hline
		& & & &\\[-6pt]  
		PSNR (dB)&32.14&32.50&32.68&32.74 \\
		\hline
		& & & &\\[-6pt]  
		MS-SSIM&0.9799&0.9832&0.9842&0.9845 \\
		\hline
	\end{tabular}
\end{table}

\begin{table}[htbp]
	\centering
	\caption{Sensitive analysis of weight parameter $\lambda_i$.}
	\label{table3}
	
	\begin{tabular}{|c|c|c|c|c|}  
		\hline 
		& & & &\\[-6pt] 
		$\lambda_i$&0.01&0.3&0.7&0.99 \\
		\hline
		& & & &\\[-6pt]  
		PSNR (dB)&32.33&32.44&32.68&32.46 \\
		\hline
		& & & &\\[-6pt]  
		MS-SSIM&0.9817&0.9824&0.9842&0.9825 \\
		\hline
	\end{tabular}
\end{table}

\begin{figure*}[htbp]
	\centering
	\includegraphics[width=6.8in]{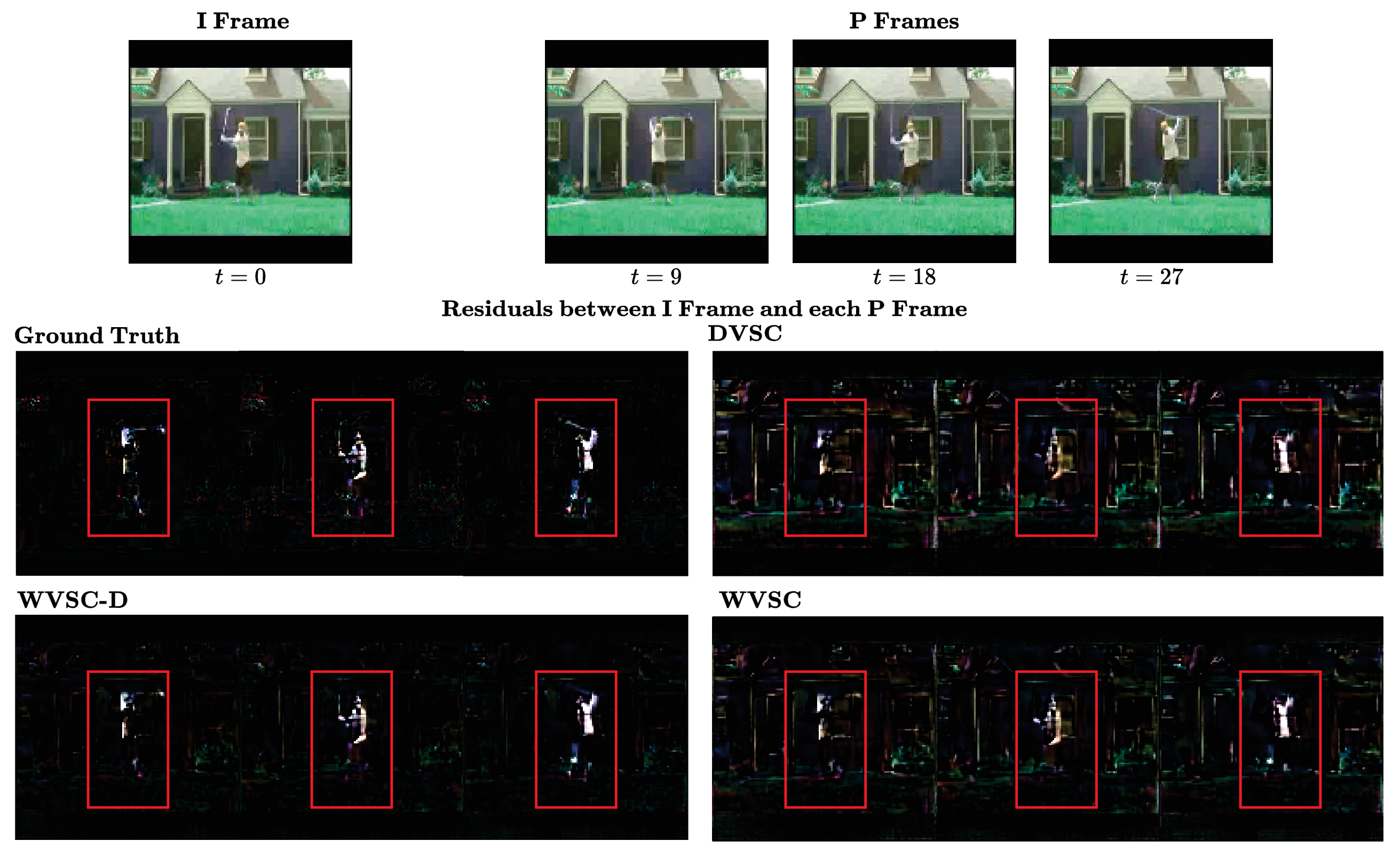}
	\caption{Visualized illustration for the effectiveness of DDMFC. Selected frame intervals with $t = 9, 18, 27$ are provided. Residuals between I frame and each P frame are presented, including the ground truth, WVSC-D, WVSC and DVSC.}
	\label{fig_11}
\end{figure*}

\subsubsection{Illustration for the Decoupled Diffusion Multi-frame Compensation}
To further illustrate the effectiveness of proposed DDMFC, we analyze the performance of DDMFC compared to other structures with frame compensation at the receiver. As shown in Fig. \ref{fig_11}, we provide the residuals between I frame and each P frame in the same GoP with different frame intervals, respectively. For the DVSC, the destruction of I frame leads to the performance degradation since the temporal information at the receiver turns to be inaccurate. Moreover, with the increase of frame intervals, it seems to be more difficult for DVSC to learn the motion differences between reference semantic frame and each current semantic frame. For the WVSC, though it adapts cross attention-based multi-frame compensation module to compensate the temporal information, the motion differences between the semantic I frame and each semantic P frame still exist. This is due to the insufficient exploration of embedded temporal information inside the previously reconstructed semantic frames. The simple fusion structure and strategy are unable to fully remove the motion differences. For the WVSC-D, DDMFC achieves much closed performances to the ground truth compared to WVSC and DVSC. Such results demonstrate that DDMFC is robust for the frame compensation task.

\subsubsection{Sensitive Analysis of different GoP Sizes}
\begin{figure}[htbp]
	\centering
	\includegraphics[width=3.3in]{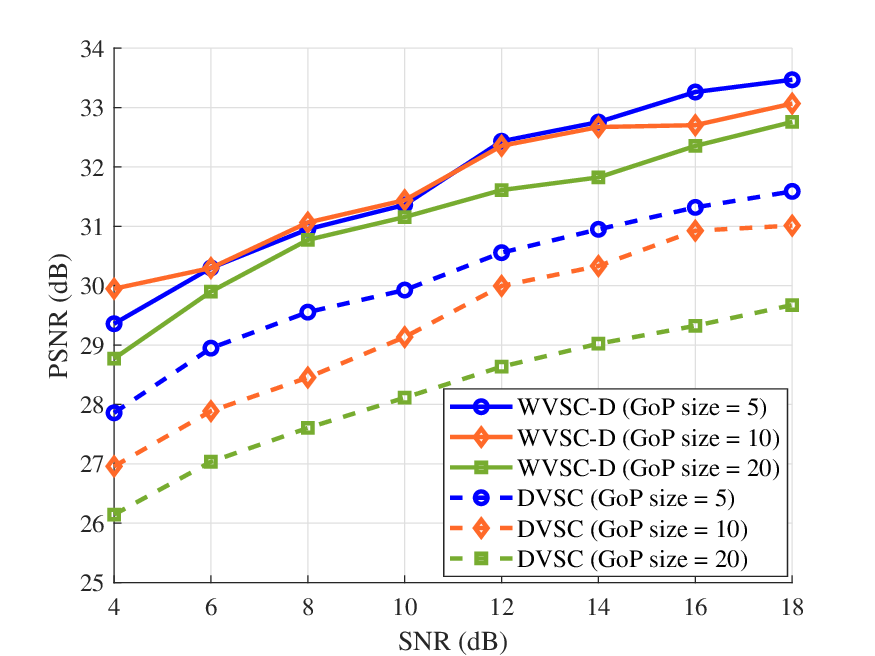}
	\caption{Performance of different GoP sizes. (all trained with GoP size 5)}
	\label{fig_10}
\end{figure}

When utilizing the first frame in a GoP as reference for transmission and video coding, a performance degradation could emerge when the GoP size becomes large. As GoP size grows, motion differences between the semantic I frame and the isolated semantic P frame increases simultaneously, bringing difficulty for the decoder to acquire motion information. As shown in Fig. \ref{fig_10}, the pretrained framework with GoP size 5 is directly employed to be tested with different GoP sizes. For WVSC-D, with the multi-frame compensation module empowered by the DDMFC, transmitting only the reference frame is enough to achieve satisfying transmission performance, as GoP size 5 and 10 do. With a large GoP size as 20, the overall performance drops but still retains relatively satisfying. The robustness of WVSC-D with different GoP sizes lies in the strong generation ability of DDMFC. With previously reconstructed frames as supplement for compensating motion information, the semantic I frame can be steered to the direction of each semantic P frame with both uniqueness and consistency. Compared to the DVSC, the performances degrade with the increase of GoP sizes. It is due to the intrinsic defect of pixel-level wireless video transmission, which depends much on the accurate motion vectors for video reconstruction. With large GoP sizes, DVSC trained with GoP size 5 is unable to match the motion differences between the latter adjacent frames. To conclude, when specifically trained with a larger GoP size, we believe WVSC could retain a satisfying performance. 

\begin{table}[htbp]
	\centering
	\caption{Evaluation of complexity and computation cost.}
	\label{table4}
	
	\begin{tabular}{|c|c|c|c|}  
		\hline 
		& & &\\[-6pt] 
		Metric&FLOPs (G)&Runtime (ms)&Parameters (M) \\
		\hline
		& & &\\[-6pt]  
		WVSC-D&109.8&40.1&39.4 \\
		\hline
		& & &\\[-6pt]  
		WVSC-G&106.3&38.4&37.6 \\
		\hline
		& & &\\[-6pt] 
		DVSC &73.6&32.8&31.6\\
		\hline
	\end{tabular}
\end{table}

\subsubsection{Complexity Analysis}
Finally, to evaluate the feasibility of WVSC-D in practical deployment, we analyze the complexity and computation cost of proposed WVSC-D and other benchmarks. FLOPs represents the number of floating-point operations that can be performed per second, which measures the computational complexity of each model. Runtime means the inference speed for transmitting videos, which refers to the runtime for transmitting a single video frame. Parameters is the total parameters of a model, which is mainly related to framework complexity. As shown in Tab. \ref{table4}, with similar Parameters, WVSC-D shows similar performance compared to WVSC-G in terms of FLOPs and Runtime along with better transmission results. This is due to the lightweight structure of U-net for residual noise generation. While for DVSC, although it has less FLOPs and higher inference speed than WVSC-D, the video reconstruction performances are much worse than WVSC-D, which turns to be a trade-off between performance gain and model complexity. Moreover, DVSC relies on the frame-by-frame reconstruction since it depends on the adjacent previous frame as reference for motion estimation and compensation. While for the WVSC-D, reference frame is assumed as the first frame in a GoP. In this way, the frame decoding stage can be computed in a parallel-efficient manner when the multi-frame conditions are acquired through emulation at the receiver during model inference. When fully being trained and deployed in practice, WVSC-D achieves satisfying Runtime performance and reasonable model complexity.

\section{Conclusion}
In this paper, we have proposed WVSC-D, which integrates semantic communication into the deep video coding designs. Deep video coding is conducted in the semantic level. The DDMFC module is proposed to polish the received reference semantic frame into the reconstructed current semantic frame with the help of previous semantic frames. In this way, both the consistency and the uniqueness for the compensated semantic frames can be improved. The superior performances of WVSC-D demonstrate the effectiveness of conducting video coding based on extracted semantics. It can be deployed in common wireless video transmission scenarios, such as point-to-point video links and Internet of Things networks. Future work will extend the framework to transmit multiple data modalities including text, audio, and images by leveraging advanced multi-modal fusion models.

\fontsize{8pt}{10pt}\selectfont

\begin{thebibliography}{50}
\bibitem{wvsc}
B. Xie, Y. Wu, Y. Shi, B. Feng, W. Zhang, J. Park, and T. Quek, ``WVSC: Wireless Video Semantic Communication with Multi-frame Compensation," \emph{arxiv:2503.21197}, Mar. 2025. [Online]. Available: \url{https://arxiv.org/abs/2503.21197}.

\bibitem{264} 
H. Schwarz, D. Marpe, and T. Wiegand, ``Overview of the H. 264/AVC video coding standard," \emph{IEEE Trans. Circuits Syst. Video Technol.}, vol. 17, no. 9, pp. 1103-1120, Sept. 2007.

\bibitem{265} 
G. Sullivan, J. Ohm, W. Han, and T. Wiegand, ``Overview of the high efficiency video coding (HEVC) standard," \emph{IEEE Trans. Circuits Syst. Video Technol.}, vol. 22, no. 12, pp. 1649-1668, Dec. 2012.

\bibitem{jscc} 
D. Gündüz, M. Wigger, T. Tung, P. Zhang, and Y. Xiao, ``Joint Source–Channel Coding: Fundamentals and Recent Progress in Practical Designs," \emph{Proc. IEEE.}, (early access), Nov. 2024. 

\bibitem{cmts} 
X. Wei, J. Liao, L. Zhou, H. Sari, W. Zhuang, ``Toward Generic Cross-Modal Transmission Strategy," \emph{IEEE Trans. Commun.}, vol. 72, no. 10, pp. 6059-6072, 2024.

\bibitem{llcm} 
Y. Chen, P. Li, A. Li, D. Wu, L. Zhou, Y. Qian, ``Toward Low-Latency Cross-Modal Communication: A Flexible Prediction Scheme," \emph{IEEE Trans. Mob. Comput.}, vol. 23, no. 12, pp. 13310-13324, 2024.

\bibitem{DeepSC}
H.~Xie, Z.~Qin, G.~Y. Li, and B.-H. Juang, ``Deep learning enabled semantic communication systems'', \emph{IEEE Trans. Signal Process.}, vol.~69, pp. 2663-2675, Apr. 2021.

\bibitem{NTSCC}
J. Dai, S. Wang, K. Tan, Z. Si, X. Qin, K. Niu, and P. Zhang, ``Nonlinear transform source-channel coding for semantic communications", \emph{IEEE J. Select. Areas Commun.}, vol. 40, no. 8, pp. 2300-2316, Aug. 2022.

\bibitem{LCFSC}
B.~Xie, Y.~Wu, Y.~Shi, W.~Z, S.~Cui, and M.~Debbah, ``Robust image semantic coding with learnable CSI fusion masking over MIMO fading channels," \emph{IEEE Trans. Wireless Commun.}, vol. 23, no. 10, pp. 14155-14170, Oct. 2024.

\bibitem{robust}
Q.~Hu, G.~Zhang, Z.~Qin, Y. Cai, G. Yu, and G. Li, ``Robust semantic communications with masked VQ-VAE enabled codebook," \emph{IEEE Trans. Wireless Commun.}, vol. 22, no. 12, pp. 8707-8722, Dec. 2023.



\bibitem{dvc} 
P. Jiang, C. Wen, S. Jin, and G. Li, ``Wireless semantic communications for video conferencing," \emph{IEEE J. Select. Areas Commun.}, vol. 41, no. 1, pp. 230-244, Jan. 2023.

\bibitem{dvsc}
H. Niu, L. Wang, Z. Lu, K. Du, and X. Wen, ``Deep learning enabled video semantic transmission against multi-dimensional noise," in \emph{Proc. IEEE Glob. Commun. Conf.  Workshops (GLOBECOM Workshops)}, Kuala Lumpur, Malaysia, pp. 1267-1272, Dec. 2023.

\bibitem{DVC} 
G. Lu, W. Ouyang, D. Xu, X. Zhang, C. Cai, Z. Gao, ``DVC: an end-to-end deep video compression framework," in \emph{Proc. IEEE Conf. Comput. Vis. Pattern Recognit. (CVPR)}, Long Beach, California, USA, pp. 11006-11015, Jun., 2019.

\bibitem{MLVC}
J. Lin, D. Liu, H. Li, and F. Wu, ``M-LVC: multiple frames prediction for learned video compression," in \emph{Proc. IEEE Conf. Comput. Vis. Pattern Recognit. (CVPR)}, Seattle, Washington, USA, pp. 3546-3554, Jun. 2020.

\bibitem{ALVC}
R. Yang, R. Timofte, and L. Van Gool, ``Advancing learned video compression with in-loosemantic P frame prediction," \emph{IEEE Trans. Circuits Syst. Video Technol.}, vol. 33, no. 5, pp. 2410-2423, May 2023.

\bibitem{fvc} 
Z. Hu, G. Lu, and D. Xu, ``FVC: a new framework towards deep video compression in feature space," in \emph{Proc. IEEE Conf. Comput. Vis. Pattern Recognit. (CVPR)}, Nashville, TN, USA, pp. 1502-1511, Jun. 2021.

\bibitem{gfvc} 
B. Chen, J. Chen, S. Wang and Y. Ye, ``Generative face video coding techniques and standardization efforts: A review," in \emph{IEEE Data Comp. Conf. (DCC)}, pp. 103-112, Mar. 2024.

\bibitem{cft} 
H. Wang, H. Li, M. Sheng, and J. Li, ``Collaborative fine-tuning of mobile AIGC models with wireless channel conditions," \emph{IEEE Wireless Commun.}, vol. 31, no. 4, pp. 32-38, Aug. 2024.

\bibitem{cdd}
H. Du, R. Zhang, D. Niyato, J. Kang, Z. Xiong, and D. Kim, ``Exploring collaborative distributed diffusion-based AI-generated content (AIGC) in wireless networks," \emph{IEEE Netw.}, vol. 38, no. 3, pp. 178-186, May 2024.

\bibitem{360dvd}
Q. Wang, W. Li, C. Mou, X. Cheng, and J. Zhang, ``360DVD: controllable panorama video generation with 360-degree video diffusion model," in \emph{Proc. IEEE Conf. Comput. Vis. Pattern Recognit. (CVPR)}, Seattle, WA, USA, Jun. 2024.

\bibitem{diffeditor}
C. Mou, X. Wang, J. Song, Y. Shan, and J. Zhang, ``DiffEditor: boosting accuracy and flexibility on diffusion-based image editing," in \emph{Proc. IEEE Conf. Comput. Vis. Pattern Recognit. (CVPR)}, Seattle, WA, USA, Jun. 2024.

\bibitem{vit}
A.~Dosovitskiy, L.~Beyer, A.~Kolesnikov, D.~Weissenborn, X.~Zhai,
T.~Unterthiner, M.~Dehghani, M.~Minderer, G.~Heigold, and S.~Gelly,
``An image is worth 16x16 words: Transformers for image recognition at
scale," in \emph{Proc. Int. Conf. Learn. Represent. (ICLR)}, Vienna, Austria, Jan. 2021. 

\bibitem{cnn}
A.~Krizhevsky, I.~Sutskever, and G.~E. Hinton, ``ImageNet classification with
	deep convolutional neural networks," in \emph{Commun. ACM}, vol.~60, no.~6,
pp. 84--90, 2017.

\bibitem{gan}
I. Goodfellow, J. Pouget-Abadie, M. Mirza, B. Xu, D. Warde-Farley, S. Ozair, A. Courville, and Y. Bengio, ``Generative adversarial nets,"  in \emph{Proc. Adv. Neural Inf. Process. Syst. (NIPS)}, Montreal, QC, Canada, Dec. 2014.

\bibitem{vae}
D.~Kingma, and M.~Welling, ``Auto-encoding variational bayes", \emph{arxiv: 1312.6114}, Dec. 2013. [Online]. Available: \url{https://arxiv.org/abs/1312.6114}.

\bibitem{dmce}
Y. Zeng, X. He, X. Chen, H. Tong, Z. Yang, Y. Guo, and J. Hao, ``DMCE: diffusion model channel enhancer for multi-user semantic communication systems," in \emph{Proc. IEEE Int. Conf. Commun. (ICC)}, Denver, CO, USA, pp. 855-860, Jun. 2024.

\bibitem{cddm}
T. Wu, Z. Chen, D. He, L. Qian, Y. Xu, M. Tao, and W. Zhang, ``CDDM: channel denoising diffusion models for wireless semantic communications," \emph{IEEE Trans. Wireless Commun.}, vol. 23, no. 9, pp. 11168-11183, Sept. 2024.

\bibitem{ddpm} 
J. Ho, A. Jain, and P. Abbeel, ``Denoising diffusion probabilistic models," in \emph{Proc. Adv. Neural Inf. Process. Syst. (NIPS)}, Online, pp. 6840-6851, Dec. 2020.

\bibitem{leakyrelu} 
J. Xu, Z. Li, B. Du, M. Zhang, and J. Liu, ``Reluplex made more practical: Leaky ReLU," in \emph{IEEE Symp. Comput. Commun. (ISCC)}, Rennes, France, pp. 1-7, Jul., 2020.

\bibitem{ddim}
J. Song, C. Meng, and S. Ermon, ``Denoising diffusion implicit models," \emph{arxiv:2010.02502}, Oct. 2022. [Online]. Available: \url{https://arxiv.org/abs/2010.02502}.

\bibitem{ebm}
W. Grathwohl, C. Wang, H. Jacobsen, ``Your classifier is secretly an energy based model and you should treat it like one," \emph{arxiv:1912.03263}, Sept. 2019. [Online]. Available: \url{https://arxiv.org/abs/1912.03263}.

\bibitem{langevin}
K. Sekimoto, ``Langevin equation and thermodynamics," \emph{Prog. Theor. Phys. Supp.}, vol. 130, pp. 17-27, Jan. 1998.

\bibitem{steer}
N. Nair, et al. ``Steered diffusion: A generalized framework for plug-and-play conditional image synthesis," in \emph{Proc. IEEE Int. Conf. Comput. Vis. (ICCV)}, Paris, France, pp. 20850-20860, Oct. 2023.

\bibitem{videofusion}
Z. Luo, D. Chen, Y. Zhang, Y. Huang, L. Wang, Y. Shen, D. Zhao, J. Zhou, and T. Tan, ``Videofusion: decomposed diffusion models for high-quality video generation," \emph{arxiv:2303.08320}, Oct. 2023. [Online]. Available: \url{https://arxiv.org/abs/2303.08320}.

\bibitem{swin}
Z. Liu, Y. Lin, Y. Cao, H. Hu, Y. Wei, Z. Zhang, S. Lin, and B. Guo, ``Swin transformer: hierarchical vision transformer using shifted windows", in \emph{Proc. IEEE Conf. Comput. Vis. Pattern Recognit. (CVPR)}, Montreal, QC, Canada, pp. 9992-10002, Oct. 2021.

\bibitem{unet}
Ronneberger, Olaf, Philipp Fischer, and Thomas Brox. ``U-net: convolutional networks for biomedical image segmentation," \emph{Proc. Int. Conf. Med. Image Comput. Comput.-Assisted Intervention (MICCAI)}, Munich, Germany, Oct. 2015.

\bibitem{cross} 
R. Li, D. Gong, W. Yin, H. Chen, Y. Zhu, K. Wang, X. Chen, J. Sun, and Y. Zhang, ``Learning to fuse monocular and multi-view cues for multi-frame depth estimation in dynamic scenes," in \emph{Proc. IEEE Conf. Comput. Vis. Pattern Recognit. (CVPR)}, Vancouver, BC, Canada, pp. 21539-21548, Jun. 2023.

\bibitem{ucf}
K. Soomro, A. Zamir, and M. Shah, ``UCF101: a dataset of 101 human actions classes from videos in the wild," \emph{arxiv:1212.0402}, Dec. 2012. [Online]. Available: \url{https://arxiv.org/abs/1212.0402}.

\bibitem{witt}
K.~Yang, S.~Wang, J.~Dai, et al. ``WITT: A wireless image transmission transformer for semantic communications", in \emph{Proc. IEEE Int. Conf. Acoust. Speech Signal Process. (ICASSP)}, Rhodes Island, Greece, Jun. 2023, pp. 1-5.

\bibitem{adamw}
I. Loshchilov, and H. Frank, ``Decoupled weight decay regularization," \emph{arxiv: 1711.05101}, Jan. 2019. [Online]. Available: \url{https://arxiv.org/abs/1711.05101}.

\bibitem{mmse}
P. Monsen, ``MMSE equalization of interference on fading diversity channels," \emph{IEEE Trans. Commun.}, vol. 32, no. 1, pp. 5-12, Jan. 1984.

\bibitem{ffmpeg}
S. Tomar, ``Converting video formats with FFmpeg," \emph{Linux J.}, vol. 2006, no. 146, Jun. 2006.

\bibitem{sionna} 
H., Jakob, et al. ``Sionna: An open-source library for next-generation physical layer research," Mar. 2022. [Online]. Available: \url{https://arxiv.org/abs/2203.11854}.

\bibitem{ssim} 
Z. Wang, E. Simoncelli, and A. Bovik, ``Multiscale structural similarity for image quality assessment," in \emph{Proc. 37th Asilomar Conf. Signals, Syst. Comput.}, vol. 2, pp. 1398-1402, Nov. 2004.

\bibitem{lpips} 
R. Zhang, P. Isola, A. Efros, E. Shechtman, and O. Wang, ``The unreasonable effectiveness of deep features as a perceptual metric," in \emph{Proc. IEEE Conf. Comput. Vis. Pattern Recognit. (CVPR)}, Salt Lake City, Utah, USA, pp. 586-595, Oct. 2018.
\end{thebibliography}

\end{document}